\documentclass[useAMS,usenatbib]{mn2e}
\input psfig.tex

\usepackage{latexsym}
\usepackage{amssymb}
\usepackage{amsmath}
\usepackage{graphics}
\usepackage{graphicx}
\usepackage{natbib}
\usepackage{subfigure}

\usepackage{bm} 
\usepackage{url}

\usepackage{morefloats}

\title[Scale-free convection in stellar models]{Theory of stellar convection II: \\
  first stellar models}
\author[S. Pasetto, C. Chiosi, E. Chiosi, M. Cropper, A. Weiss]{S. Pasetto $^{1}$\thanks{E-mail:
s.pasetto@ucl.ac.uk}, C. Chiosi $^{2}$, E. Chiosi $^{3}$, M. Cropper $^1$, A. Weiss $^{4}$ \\
 $^{1}$University College London, Department of Space \& Climate Physics, Mullard Space Science Laboratory, Holmbury St. Mary, \\ \,\,\,Dorking Surrey, United Kingdom \\
 $^{2}$Department of Physics \& Astronomy, "`Galileo Galilei"', University of Padua, Vicolo dell'Osservatorio 2, Padova, Italy\\
 $^{3}$ INAF-Osservatorio Astronomico di  Padova, Vicolo dell'Osservatorio 5, 35122 Padua, Italy\\
 $^{4}$Max-Planck-Institut f\"ur Astrophysik, Karl-Schwarzschild-Str. 1, Garching bei M\"unchen, Germany\\
 }

\begin{document}

\date{Accepted: Received in original form: }

\pagerange{\pageref{firstpage}--\pageref{lastpage}} \pubyear{2015}

\maketitle

\label{firstpage}

\begin{abstract}
We present here the first stellar models  on the Hertzsprung-Russell diagram (HRD), in which  convection is treated according to the novel scale-free convection theory (SFC theory) by
\citet[][]{2014MNRAS.445.3592P}.  The aim  is to compare the results of the new theory with those from the classical, calibrated mixing-length (ML) theory to examine differences and similarities.

We integrate the  equations describing the structure of the atmosphere from the stellar surface down to a few percent of the stellar mass using both  ML theory and SFC theory.  The key temperature over pressure gradients, the energy fluxes, and the extension of the convective zones are compared in both theories.

The analysis is first made for the Sun and then extended to other stars of different mass and evolutionary stage. The results are adequate: the SFC theory yields  convective zones, temperature gradients $\nabla$ and $\nabla_e$, and energy fluxes that are very similar to those derived from the ``calibrated'' MT theory for main sequence stars.
We conclude that the old scale dependent  ML theory  can now be replaced with a self-consistent scale-free theory able to predict correct results, one which is simpler and more physically grounded than the ML theory. Fundamentally, the SFC theory offers a deeper  insight of the  underlying physics than numerical simulations.
\end{abstract}


\section{Introduction}\label{Introduction}

Convection is one of the fundamental mechanisms for carrying  energy throughout a star from the deep interiors to the outermost layers.
This may happen during the  pre-main-sequence,  Hayashi phase of stars of any mass that are fully convective, during the main sequence phase  in central cores of stars more massive than about 1.3 $\rm M_\odot$, in the main sequence phase of very low mass stars (lower that about 0.3 $\rm M_\odot$) that remain fully convective, in the central cores of all stars burning helium and in massive stars burning heavier fuels up to iron, in  intermediate convective shells of some stars,  in the outermost layers of stars of any mass where ionization of light elements occurs, in very deep convective envelopes of red giant branch (RGB) and asymptotic giant branch (AGB) stars, in white dwarfs, in the internal regions of stars in pre-supernova stages, during the collapse phase of Type II SNae, and in the carbon-deflagration stages of type Ia SNae: convection is ubiquitous in stars of any mass and evolutionary phase. Convection significantly contributes in the transport of energy across the layers of a star, from the deep interior to the surface and also substantially changes the structure of a star by mixing the material across it.

In a star, convection sets in where and when the condition $\nabla_{{\rm{rad}}} < \nabla_{{\rm{ad}}}$ is violated, where $\nabla_{{\rm{rad}}}$ and  $\nabla_{{\rm{ad}}}$  are the radiative and adiabatic logarithmic temperature gradient with respect to pressure, i.e. ${\nabla _{{\rm{rad}}}} \equiv {\left| {\frac{{d\ln T}}{{d\ln P}}} \right|_{{\rm{rad}}}}$ and  ${\nabla _{{\rm{ad}}}} \equiv {\left| {\frac{{d\ln T}}{{d\ln P}}} \right|_{{\rm{ad}}}}$ \citep[e.g.,][]{1968pss..book.....C, 2012sse..book.....K}.
While in the inner convective regions of a star the large thermal capacity of convective elements induces a temperature over pressure gradient of the medium, $\nabla  \equiv \left| {\frac{{d{\mathop{\rm ln T}\nolimits} }}{{d\ln P}}} \right|$, that is nearly adiabatic, i.e. $\nabla - \nabla_{\rm{ad}} \simeq 10^{-8} \simeq 0$,  in the outer layers both the temperature gradients of the medium and of the element ${\nabla _e} \equiv {\left| {\frac{{d{\mathop{\rm ln T}\nolimits} }}{{d\ln P}}} \right|_e}$  differ significantly from $\nabla_{\rm{ad}}$ (super-adiabaticity). Convective elements in these regions have low thermal capacity and thus the approximation $\nabla-\nabla_{\rm ad} \simeq 0$  can no longer be applied: $\nabla_e$ and $\nabla$ must be determined separately to determine the amount of energy carried by convection (and radiation) with an appropriate theory.

Despite the great importance of convection in modelling the structure and evolution of a star, a satisfactory treatment of stellar convection  is still open to debate and until now a self-consistent description of this important physical phenomenon has been missing. Review of the current state-of-art of the turbulent non-linear magnetodynamics knowledges in the stars (mostly in the sun) can be found in several books \citep[e.g.,][]{1993noma.book.....B,2006pafp.book.....S,2012sota.book.....H}. The goal has been to establish from basic physical principles a self-consistent description of convection represented by a group of equations with no ad hoc parameters.

In this context,  the  most successful theory of stellar convection in the literature is  the Mixing-Length Theory (ML theory). The ML theory stands on the works of   \citet{Prandtl}, \citet{1951ZA.....28..304B}, and \citet{1958ZA.....46..108B}. Thanks to the success it has achieved over  decades of usage,  it is considered as  the reference paradigm to which any new theory has to be compared. In the ML theory, the motion of convective elements is expressed by means of  the mean-free-path ${l_m}$ that a generic convective element is supposed to the travel inside the convectively unstable regions of a star \citep[e.g.,][]{2012sse..book.....K}. ${l_m}$ is assumed to be proportional to the natural distance scale ${h_p}$ given by the pressure stratification of the star. The proportionality factor is called the mixing-length (ML) parameter ${\Lambda _m}$, which is implicitly defined by the relation ${l_m} = {\Lambda _m}{h_p}$. Despite this parameter must be determined with external arguments as the calibration of the ML theory on observation of the Sun, the possible dependence on the star mass and evolutionary stage, i.e. on their position in the HRD, cannot be neither excluded nor assessed. The ML parameter has paramount importance in calculating the convective energy transport, and hence the radius and effective temperature which fix the motion of the stars in the HRD.

To overcome this situation, several approaches have been proposed in literature. The simplest one  is the already mentioned fit of results obtained with a  ML parameter to observations of different stars in the CMD. In alternative formulations allow the ML-parameter to change with the position on the HRD \citep[e.g.,][]{2013MNRAS.433.2893P}. This approach is an extension of the original  idea by  \citet[][]{1958ZA.....46..108B} of the fit on the Sun. Helioseismology and/or asteroseismology
\citep{1994ARA&A..32...37B,2002RvMP...74.1073C, 2013ASPC..478..101C,2013ARA&A..51..353C} and much better data on the Sun produced by Solar Heliospheric Observatory (SOHO)  offer independent ways of testing stellar convection and constraining the ML theory in turn \citep[e.g.,][]{1977ApJ...218..521U, 1993A&A...279L...1K, 1996MNRAS.279..305G}.

Recently, sophisticated  fully 3D-hydrodynamic simulations have been used to model and test convection. This approach requires large, time consuming computational facilities to  integrate the 3D-Navier-Stokes equations \citep[e.g.,][]{1999A&A...346..111L, 2015arXiv150304582S, 2007ApJ...667..448M, 1994A&A...284..105L, 1998ApJ...496..316B, 2011JPhCS.328a2003C, 2011JPhCS.328a2012C, Magic2013a, 2015A&A...573A..89M}.

The advantage here is  that, unlike in the 1D integrations, parameter-free models of convection can be used. However,  it has a very poor interpretative power: to extract a theoretical model from a simulation is not any simpler than writing a new one from scratch.

Finally, the third approach are  ML-parameter free (or scale-free) theories by construction. It is worth mentioning, a few examples as \citet[][]{1992ApJ...397..701L}, \citet[][]{1992MNRAS.255..603B} or \citet[][]{1991ApJ...370..295C} where nevertheless other free-parameters have been used instead of the mixing-length. The turbulent scale-length in \citet[][]{1991ApJ...370..295C} is the most popular case.

In a recent paper, \citet[][]{2014MNRAS.445.3592P} developed the first theory of stellar convection  that is fully self-consistent and scale-free.  In SFC theory, the convective elements can move radially and expand/contract at the same time and in addition to the buoyancy force, the inertia of the fluid displaced by the convective elements and the effect of their expansion on the buoyancy force itself are taken into account. The dynamical aspect of the problem is differently formulated than in the classical ML theory, and the resulting equations are sufficient to determine the radiative and convective fluxes together with the medium and element temperature gradients, as well as the mean  velocity and dimensions of the convective elements as  a function of the environment physics (temperature, density, chemical composition, opacity, etc.), with no need at all of the ML-parameter. \citet[][]{2014MNRAS.445.3592P} applied the new theory to the case of the external layers of the best model representing the Sun calculated with the calibrated ML theory by \cite{2008A&A...484..815B}.

In the present study, the analysis is extended first to  model atmospheres and then to exploratory stellar models calculated with the new theory. The results are  similar to those based on the calibrated classical ML theory.

The plan of the paper is the following. In Section \ref{Outerlayers} we present the schematic structure of a star, namely the outermost layers named the photosphere fixing the radius and effective temperature, the inner atmosphere, where ionization and super-adiabatic convection occur both requiring special care, and the remaining part of the star where convection is nearly adiabatic, nuclear energy generation may change the chemical composition of the stars and convection is further complicated by the presence of over/down-shooting \citep{1962ZA.....54..114B,1964ZA.....59..215H,1994sse..book.....K}. In Section \ref{Eqset} we introduce our treatment of stellar convection as presented in \cite{2014MNRAS.445.3592P}. In Section \ref{Solve} we present the solution of the stellar equation that we are going to adopt. In Section \ref{vis_a_vis} we treat the boundary condition for the convective out-layers of the stars. In Section \ref{Results00} we present some application to the first stellar model of our theory. In Section \ref{Conclusion} we comment on our results. In Appendix A we summarize the basic equations of stellar structure in the photosphere and atmosphere,  together  a few key thermodynamical quantities concerning the equation of state (EoS)  with ionization and radiation pressure. In Appendix B, first we shortly review the classical ML theory with particular attention to the one in use here and then present the key hypotheses, assumptions and results of the new SFC theory of \citet{2014MNRAS.445.3592P} with no demonstration.


\section{Schematic structure of a star}\label{Outerlayers}

Three regions can be considered in the treatment of the physical structure of a star:
\begin{enumerate}

	\item The most external layers, i.e. the photosphere described by the optical depth, the bottom of which yields the surface of the star and determines the radius $r_*$ and effective temperature $T_{{\rm{eff}}}$.

	\item The atmosphere which extends downward for about the $3-5\%$ in mass of the star, $M_*$ from the bottom of the photosphere $\frac{M_{\rm{atm}}}{{{M_*}}} \in \left[ {1.0,0.97\div0.95} \right[$ (with this notation we specifically refer to the outer layer in radius of the star, as opposite to the central part that would be indicated as, e.g., $\frac{M}{{{M_*}}} \in \left[ {0.1,0.2} \right[$). In the atmosphere the approximation of constant luminosity (i.e. without sources or sinks of energy) can be assumed, and light elements like H and He are partially ionized. Convection is far from the regime of $\nabla -\nabla_{\rm ad} \simeq 0$. In this region both the ML theory and SFC theory find their prime application.

	\item The inner regions from  $M_{\rm{atm}}$ to the center in which energy production takes place, ionization of all elements such as H, He, C, N O etc. is complete and convection becomes adiabatic. This inner region of the stars can contain a convective envelope, extension of the convective region in the atmosphere but in which convection is nearly adiabatic. The convective envelope can extend quite deeply in the star. Stars in the mass range ${M_*} \in \left[ {0.3,1.1} \right[{M_ \odot }$ or so have a radiative core on the main sequence, stars with
$M_*\leq 0.3\, M_\odot$ are fully convective during their whole live. Stars more massive than about $M_* \geq 1.1\div1.3 M_\odot$ develop a convective core from which convective overshooting can occur. Massive stars  ($M_* \geq 10 M_\odot$) may develop intermediate convective shells in the post main sequences stages. All stars have convective cores during the core- He burning phase and beyond (their occurrence depending on the star's mass).
\end{enumerate}

The notation in use and the physical description of regions (i) and (ii) are as in the G\"ottingen stellar evolution code \citet{1964ZA.....59..215H}, the ancestor of the code used by Padova group for about five decades (see also below for more details). Details are given for the physical structure and mathematical technique used to calculate the physical variables in regions (i) and (ii)  in the Appendix A. Specifically we present the basic equations for the photosphere and atmosphere, the treatment of ionization, and a few important thermodynamical quantities such as specific heat at constant pressure $c_P$, the ambient gradient $\nabla_{\rm ad}$ and thermodynamical quantity $\Gamma_1$ for a gas in presence of radiation pressure and ionization that are needed  to describe the super-adiabatic convection both in the ML and SFC theory.

Details on the main assumptions concerning the physical input of the equations describing region (iii), the treatment of convective overshooting from the core (if required) are given in  Section \ref{Results00}.

\section{The set of equation for the SFC theory}\label{Eqset}
The system of equations Eq.(60) as in \citet[][]{2014MNRAS.445.3592P}  must  be solved to determine the convective/radiative-conductive transfer of energy in the photosphere and atmosphere of a star. They are well defined equations once the quantities $\left\{ {T,\kappa ,\rho ,{\nabla _{{\text{rad}}}},{\nabla _{{\text{ad}}}},g,{c_p}} \right\}$ are considered as input and considered constant. These quantities are, respectively, the local averaged temperature of the star, interpolated opacity tables, averaged density of the star, the radiative gradient as in Eq.\eqref{radGradB}, the adiabatic gradient as in Eq.\eqref{Adiabo} in the appendix, gravity and heat capacity at constant pressure as in Eq.\eqref{Cpionized}, considered as quantities averaged over an infinitesimal region $dr$ \textit{and time} $dt$. This means that the time-scale over which these quantities vary is supposed to be much larger than the time over which the results of the integration over the time of the system of equations for convection are achieved. Under these approximations, the system of equations proposed in \citet[][]{2014MNRAS.445.3592P} is
\begin{equation}\label{MySystem}
\left\{ \begin{array}{rcl}
{\varphi _{{\rm{rad/cnd}}}} &=& \frac{{4ac}}{3}\frac{{{T^4}}}{{\kappa {h_p}\rho }}\nabla \\
{\varphi _{{\rm{rad/cnd}}}} + {\varphi _{{\rm{cnv}}}} &=& \frac{{4ac}}{3}\frac{{{T^4}}}{{\kappa {h_p}\rho }}{\nabla _{{\rm{rad}}}}\\
\frac{{{{  v}^2}}}{{{{  \xi }_e}}} &=& \frac{{\nabla  - {\nabla _e} - \frac{\varphi }{\delta }{\nabla _\mu }}}{{\frac{{3{h_p}}}{{2\delta {{{  v}}{t_0}\tau }}} + \left( {{\nabla _e} + 2\nabla  - \frac{\varphi }{{2\delta }}{\nabla _\mu }} \right)}}g\\
{\varphi _{{\rm{cnv}}}} &=& \frac{1}{2}\rho {c_p}T\left( {\nabla  - {\nabla _e}} \right)\frac{{{{  v}^2}{t_0}\tau }}{{{h_p}}}\\
\frac{{{\nabla _e} - {\nabla _{{\rm{ad}}}}}}{{\nabla  - {\nabla _e}}} &=& \frac{{4ac{T^3}}}{{\kappa {\rho ^2}{c_p}}}\frac{{{t_0}\tau }}{{  \xi _e^2}}\\
{{  \xi }_e} &=& {\left( {\frac{{{t_0}}}{2}} \right)^2}\frac{{\nabla  - {\nabla _e} - \frac{\varphi }{\delta }{\nabla _\mu }}}{{\frac{{3{h_p}}}{{2\delta {{{  v}}{t_0}\tau }}} + \left( {{\nabla _e} + 2\nabla  - \frac{\varphi }{{2\delta }}{\nabla _\mu }} \right)}}g  \chi \left( \tau  \right),
\end{array} \right.
\end{equation}
 where ${\varphi _{{\text{rad/cnd}}}}$  is the radiative/conductive flux, $a$ the density-radiation constant, $c$  the speed of light, $T$  the local average temperature, $\kappa $  the opacity, ${h_P}$ the pressure scale height, $\rho $  the star density, ${\varphi _{{\text{cnv}}}}$ the convective flux, ${\xi _e}$ the average size of the convective cell moving with a average velocity $v$, $g$  the gravity, $\tau  = \frac{t}{{{t_0}}}$  a normalized time and $\chi  = \frac{{{\xi _e}}}{{{\xi _0}}}$  a normalized size of the convective elements. All these quantities are here treated as locally and temporally averaged.
More details on  the physical meaning of all the quantities are given in the Appendix B where a somehow different derivation of exactly these equations is explored and commented.
The form taken by the above equations in the case of  chemically homogeneous layers is straightforwardly derived from setting   $\nabla _\mu = 0 $. Let us briefly comment each equation of Eq.\eqref{MySystem} from (i) to (vi), highlighting the points of novelty with respect to the ML theory.

(i) and (ii)  In this set of equations, the first two represent  the radiative  plus conductive fluxes ${\varphi _{{\rm{rad|cond}}}}$, and the total flux ${\varphi _{{\rm{rad|cond}}}} + {\varphi _{{\rm{conv}}}}$ which defines the fictitious radiative gradient $\nabla_{{\rm{rad}}}$.

(iii) The third equation introduces one of the new aspects of the theory: the average velocity of the convective elements at a given location within the stars. Compared to the ML theory the velocity is derived from derived from the  acceleration which in turn contains  the inertia of the displaced fluid. The remarkable point of this equation is that for chemically homogeneous layers ($\nabla_{\mu}=0$) it reduces  to the equivalent in Schwarzschild approximation for stability against convection.

(iv) The fourth equation represents  the convective flux. Although the overall formulation is the same as in the ML theory, here the velocity is corrected for the effects of the inertia of the displaced fluid.  See also below for the discussion on the asymmetry of the velocity field.

(v) The fifth equation greatly differs from its analogue of the ML theory.   It measures the radiative exchange of energy between the average convective element and the surrounding medium taking into account that convective elements change their dimension, volume and area of the radiating surface as function of time because of their expansion/contraction.  In the present theory,  the energy transfer is evaluated at each instant whereas in the classical ML theory the mean size, volume and area of the emitting surface of the convective elements are kept constant.    The dependence of the energy feedback of the convective element with its surrounding is the heart of this description of  convection processes.

(vi) The last  equation yields the mean size of the convective elements as a function of time.  Its presence is particularly important because it replaces the ML theory assumption about the dimension of the convective elements and also the distance travelled by these during their lifetime. This equation achieves the closure of the system of equations.

It is worth commenting here a few aspects of the SFC theory:
\begin{itemize}
	\item \textit{Time dependence and uniqueness theorem.} The system of equations Eq.(\ref{MySystem}) and its solution contain the time. Therefore one may argue  that the mixing length, free parameter of the ML theory, is now  replaced by the time and that there is real no  advantage  with the new theory of convection but for a better description of the dynamics.  Furthermore, there are  six degrees of freedom over six unknowns instead of the five degrees of freedom over five unknowns of ML theory Eq.\eqref{MLtheoryeq1}. The solution of this apparent problem is achieved by the Uniqueness Theorem. \citet[][]{2014MNRAS.445.3592P}  have rigorously demonstrated that the ratio $  \chi / \tau^2 \to {\rm const}$ as the time grows and the solutions of the system of Eqs. (\ref{MySystem}) have to be searched in the manifold described by:
\begin{equation}\label{Eq066}
\rm \frac{ (\nabla  - \nabla _e )^2}{ \left( \nabla _{rad} - \nabla  \right) \left( \nabla_e - \nabla_{ad} \right)}{\rm{ = const}}.
\end{equation}
When the solutions enter the regime $  \chi / \tau^2 \to {\rm const}$ where the Uniqueness theorem  holds, we  simply speak of `\textit{asymptotic regime}' for the solutions. This equation  describes a surface containing the manifold of all possible solutions.  Assigned $\nabla_{\rm{rad}}$ and $\nabla_{\rm{ad}}$  that at each layer within a star, $\nabla$ and $\nabla_e $ are asymptotically related by a unique relation. There is no arbitrary scale length to be fixed. This theorem proves that, to the first order, there exists an unique manifold solution of the  system Eq.\eqref{MySystem}. The evolution of the system is forced to stay in a single time-independent manifold by the relation existing between the evolution of the average size of the convective elements and the environment where the convective elements are embedded. This relation holds only in the subsonic regime, but in such a case it is completely general\footnote{This relation can be applied to any convectively driven system suitably described by   an EoS: stars as well as planetary atmospheres, fluids, plasmas in general.}.
In our context it implies that the temporal evolution of the system Eq.\eqref{MySystem} has to cancel out: an asymptotic behaviour of the physical variables must exist. The time variable is needed to know when the asymptotic regime is reached by the system and the theory becomes fully applicable.  The Uniqueness  Theorem `de-facto' closes the equations and rules out the need of the mixing-length free parameter.

	\item \textit{Comoving reference frame $S_1$. } The advantage of an analysis made in a co-moving reference frame centred on the convective element can be captured with these simple arguments. We consider the kinetic energy associated with a convective element in the reference system ${S_0}\left( {O;{\rm{x}}} \right)$ not co-moving with the element. From the equation of the potential flow (e.g., Eq.\eqref{Eq004}), putting at rest the flow far away from the bubble (i.e. adding a flux $\left\langle {v,{\rm{x}}} \right\rangle $) we can rapidly obtain ${E_k} = \frac{1}{2}\rho \int_V^{} {{{\left\| {\rm{v}} \right\|}^2}{d^3}{\rm{x}}}  = \frac{\pi }{3}\rho \xi _e^3\left( {6\dot \xi _e^2 + {v^2}} \right)$. Here the spatial or temporal averages for the quantities ${\bar \xi _e},\bar v,...$ are omitted but they are implicitly taken into account. For the potential energy, excluding the contribution of the surface tension as mentioned above, we can write simply write ${E_p} = {E_p}\left( {{\xi _e},{P^\infty }} \right) + {\Phi _{\rm{g}}}$, i.e. the potential energy is the sum of the  potential energy of the fluid  ${E_p}\left( {{\xi _e},{P^\infty }} \right)$ that in $S_0$ is  a function of the size of the convective element and of the pressure far away from the bubble, and the gravitational potential of the star. If we limit ourselves to the equations of motion (EoM) for the radial direction outside the stars, $r$, the EoM reads:
\begin{equation}\label{LAgrange}
\begin{gathered}
  \left\{ \begin{gathered}
  \frac{d}{{dt}}\left( {\frac{{\partial L}}{{\partial {{\dot \xi }_e}}}} \right) - \frac{{\partial L}}{{\partial {\xi _e}}} = 0 \hfill \\
  \frac{d}{{dt}}\left( {\frac{{\partial L}}{{\partial \dot r}}} \right) - \frac{{\partial L}}{{\partial r}} = 0 \hfill \\
\end{gathered}  \right. \Leftrightarrow  \hfill \\
  \left\{ \begin{gathered}
  \frac{3}{2}\dot \xi _e^2 + {{\ddot \xi }_e}{\xi _e} - \frac{{P\left( {{\xi _e}} \right) - {P^\infty }}}{\rho } = 0 \hfill \\
  A = 2g, \hfill \\
\end{gathered}  \right. \hfill \\
\end{gathered}
\end{equation}
with $L$ the Lagrangian of the system.  Thanks to the lagrangian formalism, knowing  the explicit formulation for ${E_p}\left( {{\xi _e},{P^\infty }} \right)$ is not necessary, because it satisfies the equation  ${E_p}\left( {{\xi _e},{P^\infty }} \right) =  - \int_V^{} {\left( {P\left( {{\xi _e}} \right) - {P^\infty }} \right){d^3}{\rm{x}}} $. With the aid of the equation for hydrostatic equilibrium  we re-obtain Eq.\eqref{Eq010} of the first Theorem in \citet{2014MNRAS.445.3592P} but without the term containing the acceleration $A$  (or with $A = 0$) that  provides the dynamical coupling between $S_0$ and $S_1$. This forced us to apply directly the Lagrangian formalism to non-inertial reference frames as already  done by \citet[][Section 3.1]{2009A&A...499..385P}  in a different context
\citep[see also][Section 39 for the point mass approximation]{1969mech.book.....L}. Finally, we note  how this suggests also a different, completely independent derivation of the main theorem of Sec 4.1 of \citet{2014MNRAS.445.3592P} from a Lagrangian formalism.

	\item \textit{Local departure from hydrostatic equilibrium}. The SFC theory  of convection is based on the assumption of non-local pressure equilibrium and hence local deviations from rigorous hydrostatic equilibrium (a situation thereinafter shortly referred to as mechanical equilibrium),   i.e. the stellar plasma is not in mechanical equilibrium on the surface of the expanding/contracting convective element while this latter is moving outward/inward. A convective element  coming into existence for whatever reason and expanding into the medium represents  a perturbation of local pressure that cannot instantaneously recover the mechanical equilibrium (pressure balance) with the surrounding.  The condition of rigorous mechanical equilibrium with the stellar medium is met only 'far away' from the surface of a convective element, i.e. only in the limit $\xi_e \rightarrow\infty$.
	
In many textbooks of stellar astrophysics  \citep[e.g.,][]{1968pss..book.....C, 2012sse..book.....K},  the simple assumption of  instantaneous pressure equilibrium is made because of the fast expansion of the convective elements. Indicating the pressure  difference between the element (at the surface) and medium as   $DP \equiv P - {P^\infty }$, it is generally assumed $DP = 0$ identically. This is clearly a wrong assumption if one needs to argue about local deviations from hydrostatic equilibrium. There exists no instantaneous pressure equilibrium. No matter how fast it is reached, the sound speed is not infinite. We take advantage of the  Eq.\eqref{LAgrange} (which is a particular case of theorem of Sec 4.1 in \cite{2014MNRAS.445.3592P}) to prove this statement.  In Fig. \ref{Pressure}, we show the temporal behaviour of  the ratio $DP / {P^\infty }$  derived from the first Lagrangian equation of System Eq.\eqref{LAgrange} using, e.g.,  for the temporal evolution of $\xi_e$ an arbitrary relation of the type ${\xi _e} \propto {\tau ^2}$.
\begin{figure}
\centering
\resizebox{\hsize}{!}{\includegraphics{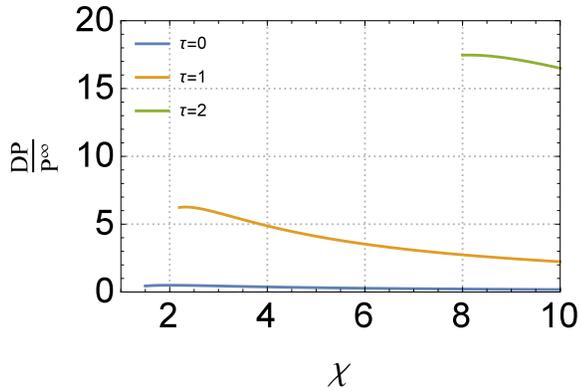}}
\caption{Evolution with the time of the pressure difference $DP/P^\infty$ at the surface of a generic convective element. The time necessary to reach the pressure equilibrium is not null, and the equilibrium  it is reached only 'far away' form the surface of the convective element. Ideally,  a convective element cannot exist at all,  if it is assumed to be always in perfect mechanical equilibrium with the environment.     }
 \label{Pressure}
\end{figure}
As evident from Fig.\ref{Pressure}, only far away from the convective element $\frac{P}{{{P^\infty }}} \to 1$, i.e., $\frac{D P}{{{P^\infty  }}} \to 0$(\footnote{Note that the convolution of the pressure profile $P = P\left( {r,t} \right)$ gives an universal profile predicted by the theory that is recovered also in our numerical models of convection in the stellar atmospheres, see Fig.\ref{profile} below. The investigation of this issue is deferred to future studies.}).
Even though the theory does not require the mechanical  equilibrium for the convective element, since the star as a whole is in hydrostatic equilibrium the condition is also formally met for the equation of  convection, but only far away from the surface of  the convective element.
However, by 'far away' we mean always a distance close enough so that the local density has remained nearly constant. This description of the physical situation as far as the mechanical equilibrium at the surface of a convective element is concerned agrees with current understanding of fluid dynamics \citep{1959flme.book.....L,2000ifd..book.....B} and the current 'gross' assumption made by the classical ML theory that convective elements expand contract in mechanical equilibrium with the surroundings.

	\item \textit{Surface tension on convective elements.} It is worth recalling attention here that in the SFC theory  no physical  surface is enclosing a convective element and therefore the Young-Laplace treatment of the surface tension is not applied. This approach differs from classical literature on fluids in which the surface tension is taken into account \citep[e.g.,][and reference therein]{Tuteja2010}. Our approach  is consistent  with astrophysical 3D-hydrodynamical simulations where convection is represented by small volumes moving up and down for a short time, not surviving enough surface tensions to be relevant.
\end{itemize}

\textsf{More general remarks}:
In addition to be above issues, we would like to shortly comment here on points of strength and weakness of the SFC theory that deserve further investigation.

Since the early studies of  Boussinesq  (1870),  Prandtl (1925) and on the Reynolds stress model, the closure of the hierarchy of averaged moment equations represented a formidable challenge for the description of turbulence and convection. Despite its simplicity, our model represents the first and to date unique way to close the equation of stellar convection, without any arbitrarily free assumptions.
It is fully analytical: neither ad hoc fitting on HRD stars nor numerical simulations are required to find closure of the equations. Furthermore, rotation can be  implemented in a simple fashion because the formalism of accelerated reference frames of type $S_1$ is already in situ. Finally, in the SFC theory the convective transport of the energy is mainly driven by the expansion of the elements and less  by their vertical motion, making simpler its inclusion with large eddies turbulence theories for high Reynolds numbers.

Up to now only few points of weakness  have been identified. First of all, the SFC theory is a linear theory. Therefore it cannot deal with non linear phenomena that would require higher order expansion over $\varepsilon \equiv \frac{|v|}{|\xi_e|} \ll 1$ and a suitable treatment of  resonances. This  problem has not been investigated yet in the context of the present theory for the stellar plasma but it may have strong physical implications. The uniqueness theorem that  provides the closure  of the stellar equation does not hold in the non linear regime. Hence, we expect that second order effects will require further free parameters, as it was the case of the ML theory, to reach a finer physical description.

Closely related to the previous problem is the determination of the distribution function of the size of the convective cells. The distribution function of a turbulent cascade of eddies is not Gaussian. In our study, we consider only with the first order moments of the -unknown- underlying distribution function. This approach is far from being a correct description of the reality. However, even if the number of underlying moments required to map correctly the distribution function and hence the nature of the convection within the stars is surely very high, it is  not infinite \citep[see, e.g.,][]{2010A&A...522A..30C}.

The size of the convection cells is expected to span from large integral scales containing the most of the kinetic energy in an anisotropic motion down to the Kolmogorov's micro-scales of the small eddies with randomly isotropic motion where the viscous forces are effective (at least for high Reynold numbers).   For the sake of simplicity,  dealing with the stellar plasmas,   we limit ourselves to a simple treatment favouring the self-consistency to the complexity\footnote{Concerning the chaotic vs. turbolent nature of the inertia term that we retained in Navier - Stokes equations see, e.g., \cite{1989kimi.book.....O}.}.

Finally, by construction, the present SFC theory does not suitably describe the border regions of convective zones  where convective cells could overshoot from the Schwarzschild border into the surrounding radiative regions. This is possible only suitably modifying the present SFC theory \citep[][in preparation]{Pasettoetal2016}.

Basing on the  results that we are going to present here,  we are confident  that some of the above criticisms will be found  to be unimportant. As a matter of fact, the first stellar  models, with the envelope convection  treated according to  the \cite{2014MNRAS.445.3592P} theory are much similar to those derived from  the best tuned ML theory. Implications of these results in relation with the above criticisms will discussed in Section\ref{Conclusion}.

\section{Solving the basic equations}\label{Solve}
We present an algebraic numerical procedure to solve  the system of Eq.\eqref{MySystem}.  Following \citet[][]{2014MNRAS.445.3592P} we  assume $ g_4 = \frac {g}{4}$ and $\alpha = \frac {a c T^3}{\kappa \rho^2 c_p} $ where all the symbols have their usual meaning and we  limit ourself to the homogeneous case ${\nabla _\mu } = 0$.  Inserting  the first of Eqs.\eqref{MySystem} into the second, and performing a number of algebraic manipulations, the system Eq. \eqref{MySystem} rapidly reduces to this set of four equations in four unknowns:
\begin{equation}\label{MySystem02}
	\left\{ \begin{array}{rcl}
	{{  v}^2} &=& \frac{{4{g_4}{{  \xi }_e}({\nabla} - {\nabla _e})}}{{\frac{{3{h_p}}}{{2\delta   v\tau }} + 2{\nabla} + {\nabla _e}}}\\
	\frac{{4\alpha ({\nabla _{{\rm{rad}}}} - {\nabla})}}{{3{h_p}}} &=& \frac{{{{  v}^2}{t_0}\tau ({\nabla} - {\nabla _e})}}{{{h_p}}}\\
	\frac{{{\nabla _e} - {\nabla _{{\rm{ad}}}}}}{{{\nabla} - {\nabla _e}}} &=& \frac{{4\alpha \tau }}{{{{  \xi }_e}^2}}\\
	{{  \xi }_e} &=& \frac{{{g_4}  \chi ({\nabla} - {\nabla _e})}}{{\frac{{3{h_p}}}{{2\delta   v\tau }} + 2{\nabla} + {\nabla _e}}}.
	\end{array} \right.
\end{equation}
There are several different techniques for finding the solutions  of the system Eq.\eqref{MySystem02}. In what follows we limit ourselves to present the most stable of these solutions from a algebraic/numerical point of view.

From the second equation of system Eq.\eqref{MySystem02} we isolate the gradient of the convective elements:
\begin{equation}
	\nabla _e= \frac{4 \alpha  (\nabla -\nabla _{\text{rad}})}{3  {v}^2 t_0 \tau }+\nabla,
\end{equation}
and insert it into the other equations of system Eq.\eqref{MySystem02} to obtain
\begin{equation}
	\left\{ \begin{array}{rcl}
	{{  v}^2} &=& \frac{{32\alpha \delta {g_4}{{  \xi }_e}({\nabla _{{\rm{rad}}}} - \nabla )}}{{2\delta \nabla \left( {4\alpha  + 9{{  v}^2}{t_0}\tau } \right) + 9  v{h_p}{t_0} - 8\alpha \delta {\nabla _{{\rm{rad}}}}}}\\
	\frac{{3{{  v}^2}{t_0}\tau ({\nabla _{{\rm{ad}}}} - \nabla )}}{{4\alpha (\nabla  - {\nabla _{{\rm{rad}}}})}} &=& \frac{{4\alpha \tau }}{{{{  \xi }_e}^2}} + 1\\
	{{  \xi }_e} &=& \frac{{8\alpha \delta {g_4}  \chi ({\nabla _{{\rm{rad}}}} - \nabla )}}{{2\delta \nabla \left( {4\alpha  + 9{{  v}^2}{t_0}\tau } \right) + 9  v{h_p}{t_0} - 8\alpha \delta {\nabla _{{\rm{rad}}}}}}.
	\end{array} \right.
\end{equation}
We proceed further by extracting the  ambient gradient from the first of the previous equations:
\begin{equation}\label{system04}
\nabla  = \frac{{8\alpha \delta {\nabla _{{\rm{rad}}}}\left( {{{  v}^2} + 4{g_4}{{  \xi }_e}} \right) - 9{{  v}^3}{h_p}{t_0}}}{{18\delta {{  v}^4}{t_0}\tau  + 8\alpha \delta \left( {{{  v}^2} + 4{g_4}{{  \xi }_e}} \right)}},
\end{equation}
and introduce it into the remaining two equations to obtain a simple equation, that relates $\xi$ and $  v$:
\begin{equation}\label{velxi1}
		4  {\xi }_e=\frac{ {v}^2  {\chi }}{ {\xi }_e},
\end{equation}
and a more complicated equation that relates all the other quantities:
\begin{equation}\label{Ancora}
	\begin{array}{l}
	- 9{{  v}^3}{h_p}{t_0} - 8\alpha \delta {\nabla _{rad}}\left( {{{  v}^2} + 4{g_4}{{  \xi }_e}} \right)\\
	+ 2\delta {\nabla _{ad}}\left( {9{{  v}^4}{t_0}\tau  + 4\alpha \left( {{{  v}^2} + 4{g_4}{{  \xi }_e}} \right)} \right)\\
	= \left( {\frac{{4\alpha \tau }}{{{{  \xi }_e}^2}} + 1} \right)\frac{{12\alpha   v(2\delta   v{\nabla _{rad}}\tau  + {h_p})}}{\tau }.
	\end{array}
\end{equation}
The first of these equations, Eq.\eqref{velxi1}, offers an immediate solution for the size  and/or  velocity of a convective element. Once Eq.\eqref{velxi1} is inserted in  Eq.\eqref{Ancora} we obtain a quintic equation in $  \xi_e$ (the current size of a convective element in $S_1$).  At this point, we are tempted to exploit the fact that by construction  $  \xi_e$  is always positive, and search for real positive solutions of the quintic equation in $  \xi_e$:
\begin{equation}\label{Quintic01}
	\begin{array}{l}
	\sum\limits_{i = 0}^5 {{c_i}  \xi _e^i}  = 0\\
	\left\{ \begin{array}{l}
	c_5 = 1\\
	c_4 = \frac{\pm h_P \sqrt{\chi }}{4 \delta  \nabla _{\text{ad}} \tau }\\
	c_3 = \frac{\alpha  t_0 \chi  (\nabla _{\text{ad}}+2 \nabla _{\text{rad}})}{9 \nabla _{\text{ad}} \tau }\\
	c_2 = \frac{12 \alpha  \pm h_P t_0 \chi ^{3/2}+\alpha  \delta  g_4 t_0{}^3 \tau  \chi ^2 (\nabla _{\text{ad}}-\nabla _{\text{rad}})}{144 \delta  \nabla _{\text{ad}} \tau ^2}\\
	c_1 = \frac{4 \alpha ^2 t_0{}^2 \nabla _{\text{rad}} \chi }{3 \nabla _{\text{ad}}}\\
	c_0 = \pm \frac{{{\alpha ^2}{h_P}t_0^2{\chi ^{3/2}}}}{{3\delta {\nabla _{{\text{ad}}}}\tau }},
	\end{array} \right.
	\end{array}
\end{equation}
where ${c_i} \in \mathbb{R}_0^ +$, $i=1,...,5$ and ${  \xi _e} \in \mathbb{R}_0^ + $. Nevertheless this apparent advantage is not so helpful in the practice. The solution of a quintic equation represents a formidable problem that kept occupied the most eminent minds of the past centuries and only in the $19^{th}$ century a solution formula in terms of ultra-radicals (elliptic functions) has been found  \citep[connection with icosahedral symmetry,][]{KingAffiliated}. The implementation of this technique, despite offering a general analysis of Eq.\eqref{Quintic01}, is beyond the goal of the present paper. We are mostly focusing on the impact and physical meaning of the convection Eq.\eqref{MySystem02} for the stars and on validating our theory. To this aim, we  make use of physical assumptions and the theorem of uniqueness  in \citet{2014MNRAS.445.3592P}    to  reach a comprehensive interpretation of our system. Hence, we omit to develop a complete mathematical treatment of the quintic Eq.\eqref{Quintic01} and proceed with the following arguments.

The average size of the convective elements  is in  bijective relation with the time (see \cite{2014MNRAS.445.3592P} Appendix A, Fig. A1). Our theory is valid only after some  time interval has elapsed since the birth of a convective element (the time interval is however small compared to any typical evolutionary time-scale of a star). Similar considerations apply to the size of a convective element.  Therefore both $\tau$ and $  \xi_e$ represent equally useful (unbounded) independent variables over which to solve our equations. The theorem of uniqueness proved that the system Eq.\eqref{MySystem} has to develop an asymptotic behaviour for the independent variables and hence, e.g., on the velocity $  v$ too.

Because a quintic equation has solutions only in terms of ultra-radicals, we find more convenient first
to express the quintic equation in terms of $  v$ and then to operate numerically to solve it.
The advantage and simplicity in determining numerically an asymptotic velocity $  v$ overwhelms in the practice the utility of the positive nature of $  \xi_e$.  Hence, we here propose to replace in  Eq. \eqref{velxi1} the variable  $  \xi_e$  with the variable $  v$ to obtain the quintic as function of $  v$ (where positive and negative solutions have to be investigated).
From Eq.\eqref{velxi1} we get ($ {\chi }>0$ and ${  \xi _e} > 0$, see also Appendix B):
\begin{equation}
	{  \xi _e} = \frac{{\left| {  v} \right|\sqrt \chi  }}{2},
\end{equation}
so that the system Eq.\eqref{MySystem02} reduces to:
\begin{equation}\label{Quintic02}
	\begin{array}{l}
		\sum\limits_{i = 0}^5 {{c_i}  v^i}  = 0\\
	\left\{ \begin{array}{l}
  {c_5} = 1 \\
  {c_4} = \frac{h_P}{2 \delta  t_0 \nabla _{\text{ad}} \tau } \\
  {c_3} = \frac{4 \alpha  (\nabla _{\text{ad}}+2 \nabla _{\text{rad}})}{9 t_0 \nabla _{\text{ad}} \tau } \\
  {c_2} = -\frac{\alpha  \left(\pm \delta  g_4 t_0{}^2 \tau  \sqrt{\chi } (\nabla _{\text{rad}}-\nabla _{\text{ad}})-12 h_P\right)}{18 \delta  t_0{}^2 \nabla _{\text{ad}} \tau ^2} \\
  {c_1} = \frac{64 \alpha ^2 \nabla _{\text{rad}}}{3 t_0{}^2 \nabla _{\text{ad}} \chi }\\
  {c_0} = \frac{32 \alpha ^2 h_P}{3 \delta  t_0{}^3 \nabla _{\text{ad}} \tau  \chi }.
	\end{array} \right.
	\end{array}
\end{equation}
Owing to the odd velocity dependence of Eq. (\ref{Quintic02}), the theory predicts different average velocities for up/down motions of convective elements. This effect has been already observed in numerical simulations \citep[e.g.,][]{2015ApJ...809...30A} and here now evident  in our fully analytical treatment.
The implication of this effect will be examined
in  greater detail in a forthcoming paper \citep[][in preparation]{Pasettoetal2016} where over/under-shooting of the convective elements will be investigated.

Therefore, at each layer of the convective regions (which means at assigned input physics: density, temperature, etc.) the time has to be changed until the ``so-called'' asymptotic regime is reached (see Section \ref{Eqset}). The time scanning is made according to the relation $t = 10^{e+\Delta e}$ [s] where, e.g., $e = 1,2,...,15$ in steps of $\Delta e$. $\Delta e$ is suitably chosen according to the desired time space and accuracy. Typical values $\Delta e \in \left[ {0.01,0.05} \right]$ produce fine resolution for the purpose of our work. At each time step $\hat t$ the integration of  the quintic of Eq.\eqref{Quintic02} is performed with robust numerical algorithm \citep[][]{1970JT} and with the solution meeting the conditions
\begin{equation}\label{Conds01}
	\left\{ \begin{gathered}
		\operatorname{Im} \left[ {  v} \right] = 0 \hfill \\
		{{  \xi }_e}\left( {\hat t} \right) > {{  \xi }_e}\left( {\hat t - dt} \right) \hfill \\
		  v\left( {\hat t} \right) >   v\left( {\hat t - dt} \right) \hfill \\
		\frac{{  v\left( {\hat t - dt} \right)}}{{  v\left( {\hat t} \right)}} > \Pi \, ,  \hfill \\
	\end{gathered}  \right.
\end{equation}
where $\Pi$ is a suitable percentage of the asymptote reached, for instance 98\% (i.e. $\Pi = 0.98$ in our notation). When this occurs, the velocity has reached its asymptotic value and the solutions are determined.
\begin{figure*}
\centering
\resizebox{\hsize}{!}{\includegraphics{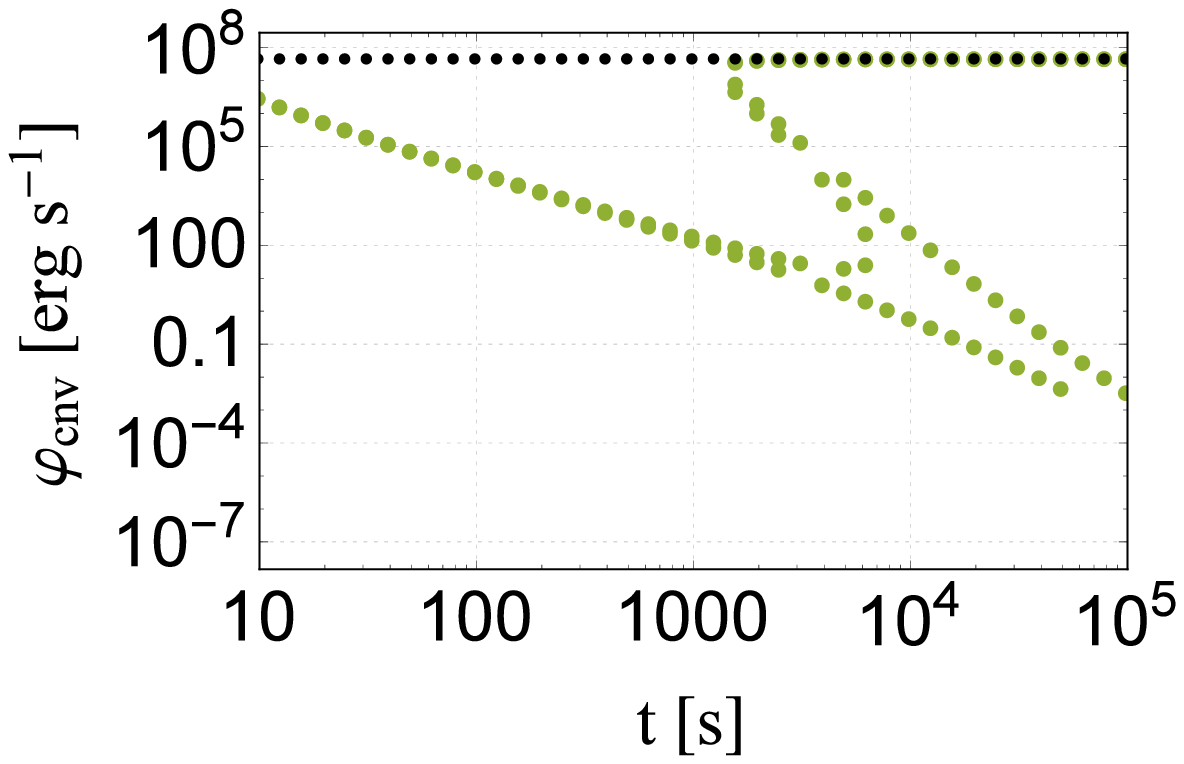}\includegraphics{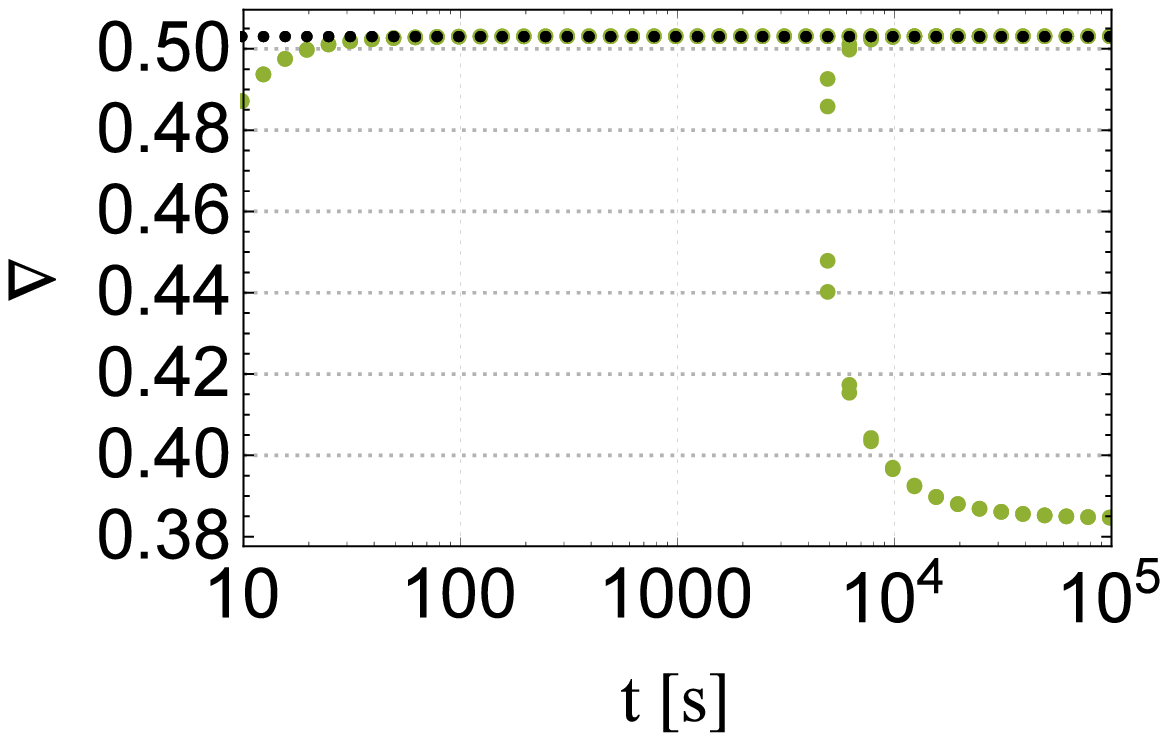}}
\caption{
Convective Flux (left panel) and temperature gradient $\nabla$  of the ambient medium (right panel) as a function of time for a layer in the outer convective zone of the main sequence model of the $1\,  M_\odot$ star with chemical composition $[X=0.703,\, Y = 0.280,\, Z = 0.017]$. In the two panels, we show the solutions from the quintic Eq.\eqref{Quintic02} of  the SFC theory with their multiplicity (green dots) and the corresponding ones from the ML theory (black dots) with parameter $\Lambda_m=1.68$. For times longer than a few $10^4$ s one of the solutions from the SFC theory almost coincides with that from the ML theory. This layer falls at the inner edge of the region with strong super-adiabaticity.}
 \label{soloplots}
\end{figure*}
\begin{figure*}
\centering
\resizebox{\hsize}{!}{\includegraphics{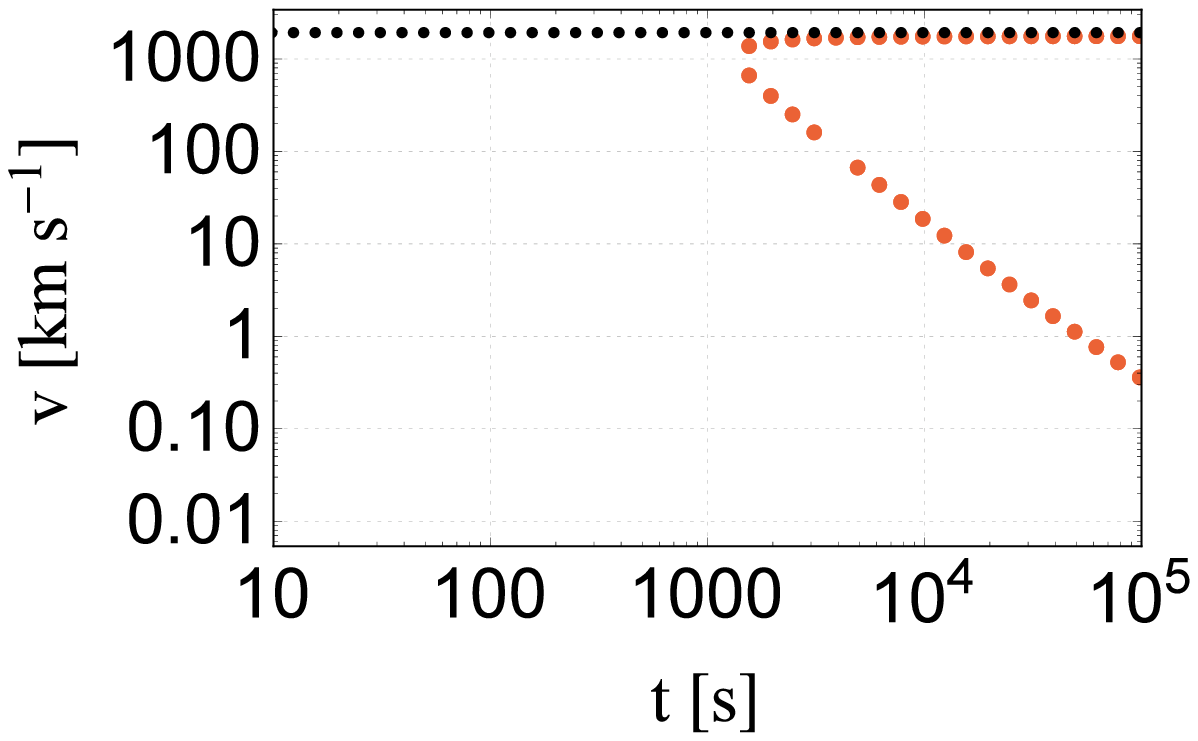}\includegraphics{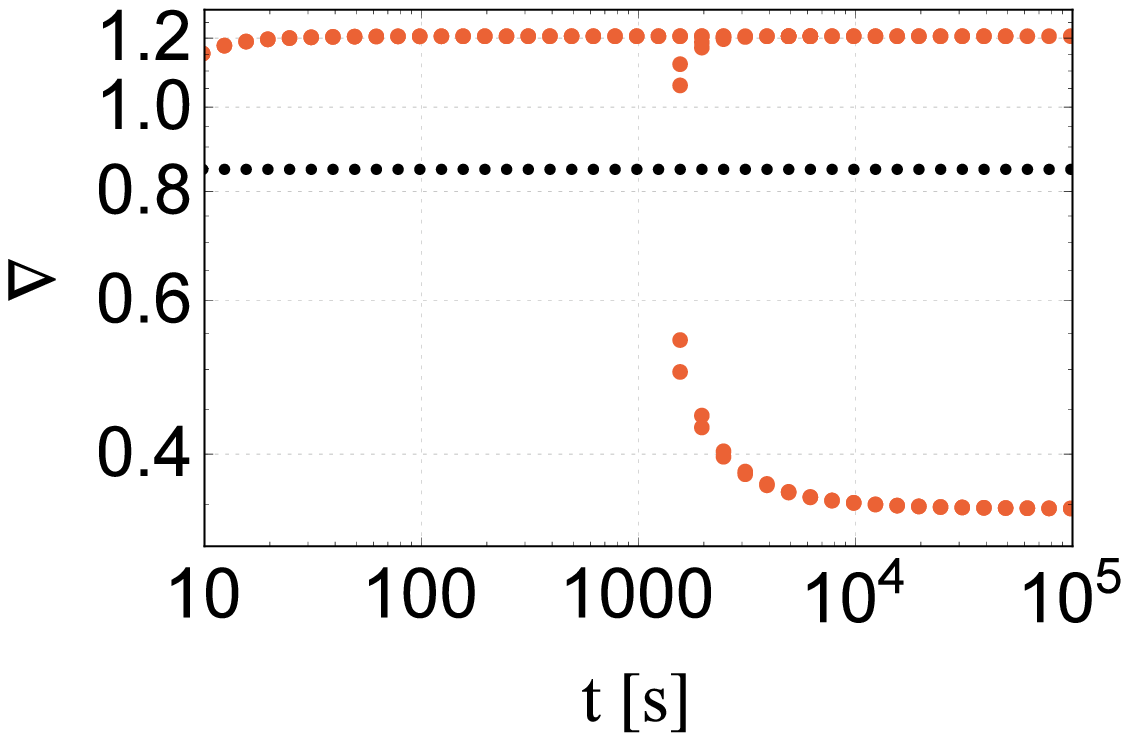}}
\caption{Similar to Fig.\ref{soloplots} but for the velocity (left panel) and the ambient temperature gradient $\nabla$ (right panel) and for a different layer of the outer convective zone of the $1\, M_\odot$ with composition $[X=0.703,\, Y = 0.280,\, Z = 0.017]$ on the main sequence. This layer is below the photosphere but above the region of very strong super-adiabaticity. The physical units and the meaning of the symbols are the same as in Fig. \ref{soloplots}. The degree of super-adiabaticity is larger that in Fig. \ref{soloplots}. While the large values of the velocities from the SFC theory  coincide with those from the  ML theory for times longer that  few $10^3$ s, the temperature gradient does not.  }
 \label{supplots}
\end{figure*}
When  the velocity $  v$ of a typical convective element is known, one can immediately calculate its  dimension
$ \xi_e$ and temperature gradient $\nabla_e$, the temperature gradient of the medium $\nabla$, the convective flux $\varphi_{\rm{cnv}}$, and finally the radiative flux $\varphi_{\rm{rad|cnd}}$.  As the quintic equation contains the integration time $\tau$,  all these  quantities vary with time until they reach their asymptotic value. Furthermore, at each time the quintic equation has solutions of which only those with null imaginary part have physical meaning and only those satisfying all the selection criteria Eq.(\ref{Conds01}) have to be considered.  To illustrate the point, we take a certain layer  located somewhat inside  the external convective zone. The layer is at the inner edge of the super-adiabatic zone  and it characterised by  the following values of the physical quantities $R=6.06736\times 10^8$ m, $T = 6.29506\times 10^3$ K, $P =1.48594\times 10^4 \,\rm Newton\, m^{-2}$, $\rho  = 3.63078\times 10^{-4} \rm \, kg \, m^{-3}$, $\kappa  = 9.67164\times 10^{-2} \, \rm m^2\, kg^{-1}$, $\nabla_{{\rm{ad}}}= 0.384$, $\nabla_{{\rm{rad}}}  = 0.503$, and $\mu = 1.279$,  solve the quintic equation and derive the whole set of unknowns listed above as a function of time until they reach the asymptotic value. The results are shown in
Fig.\ref{soloplots} limited to the convective flux $\varphi_{\rm{cnv}}$ (left panel)  and ambient temperature gradient $\nabla$ (right panel). In this Figure we display all the physical solutions, i.e. with $\rm{Im}[v]=0$. These are indicated by the green dots. Looking at the left panel, at increasing time the number of real solutions varies from one to five and past $10^4$ s to three and asymptotically only two. Similar trend is shown by the plot in the right panel.  The same quantities can also be obtained from the  classical ML theory using the equations presented in Section \ref{ML theory}. In this case only one real solution exists at each time. This is indicated by the  black dots in both panels. Finally, of all the solutions given by the SFC theory only one is filtered by the selection criteria, i.e the one with the highest value of $\varphi_{\rm{cnv}}$  and $\nabla$. Therefore, in this layer the asymptotic value of the SFC theory solution is the same as that of the ML theory.

Is this situation the same for all layers of the convective zone? It answer is no, the SFC theory differs from the ML theory in the outermost regions, whereas it closely resembles the ML theory going deeper and deeper inside. The issue is examined in detail below.
\begin{figure}
\centering
\resizebox{\hsize}{!}{\includegraphics{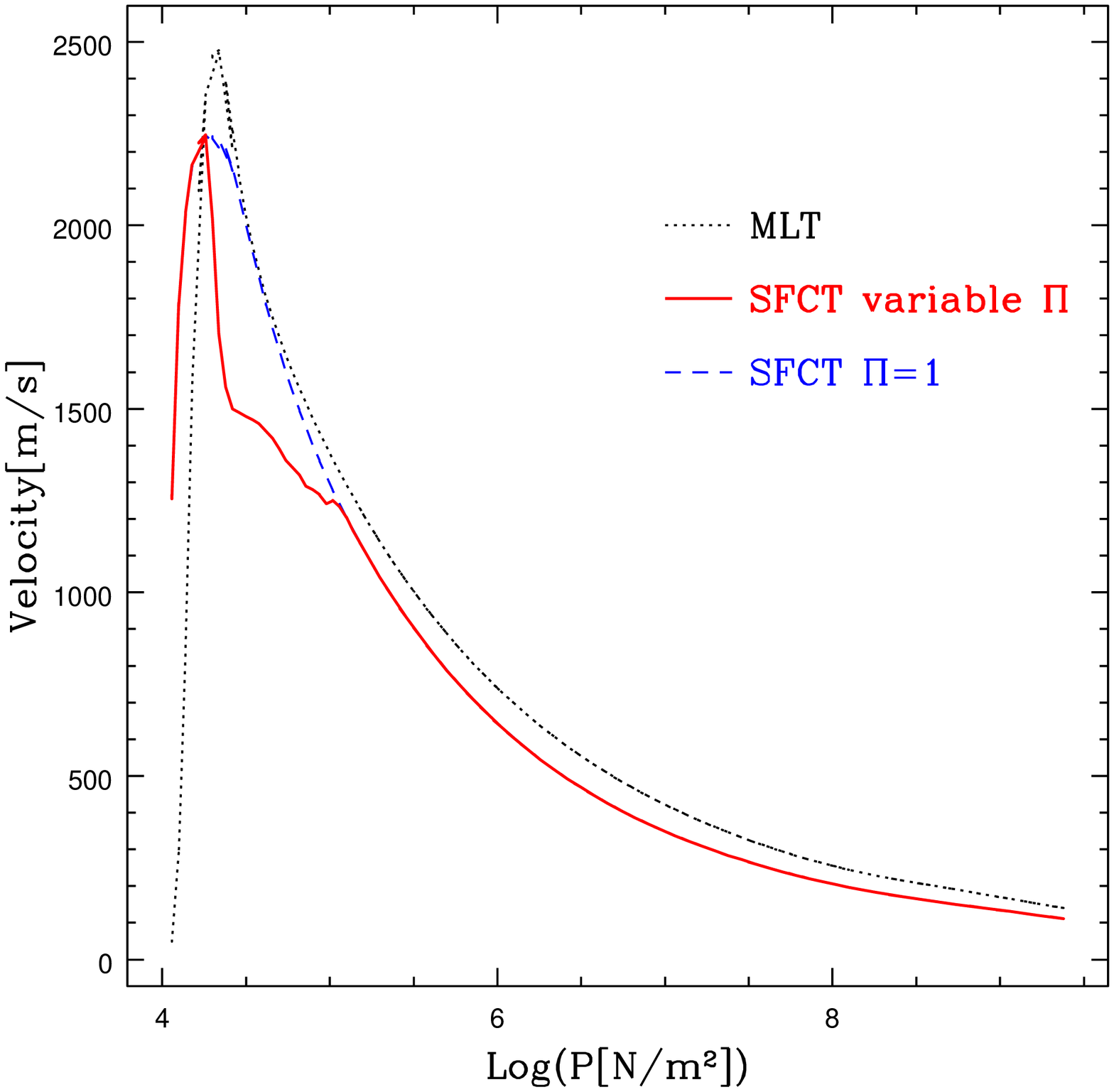} }
\resizebox{\hsize}{!}{\includegraphics{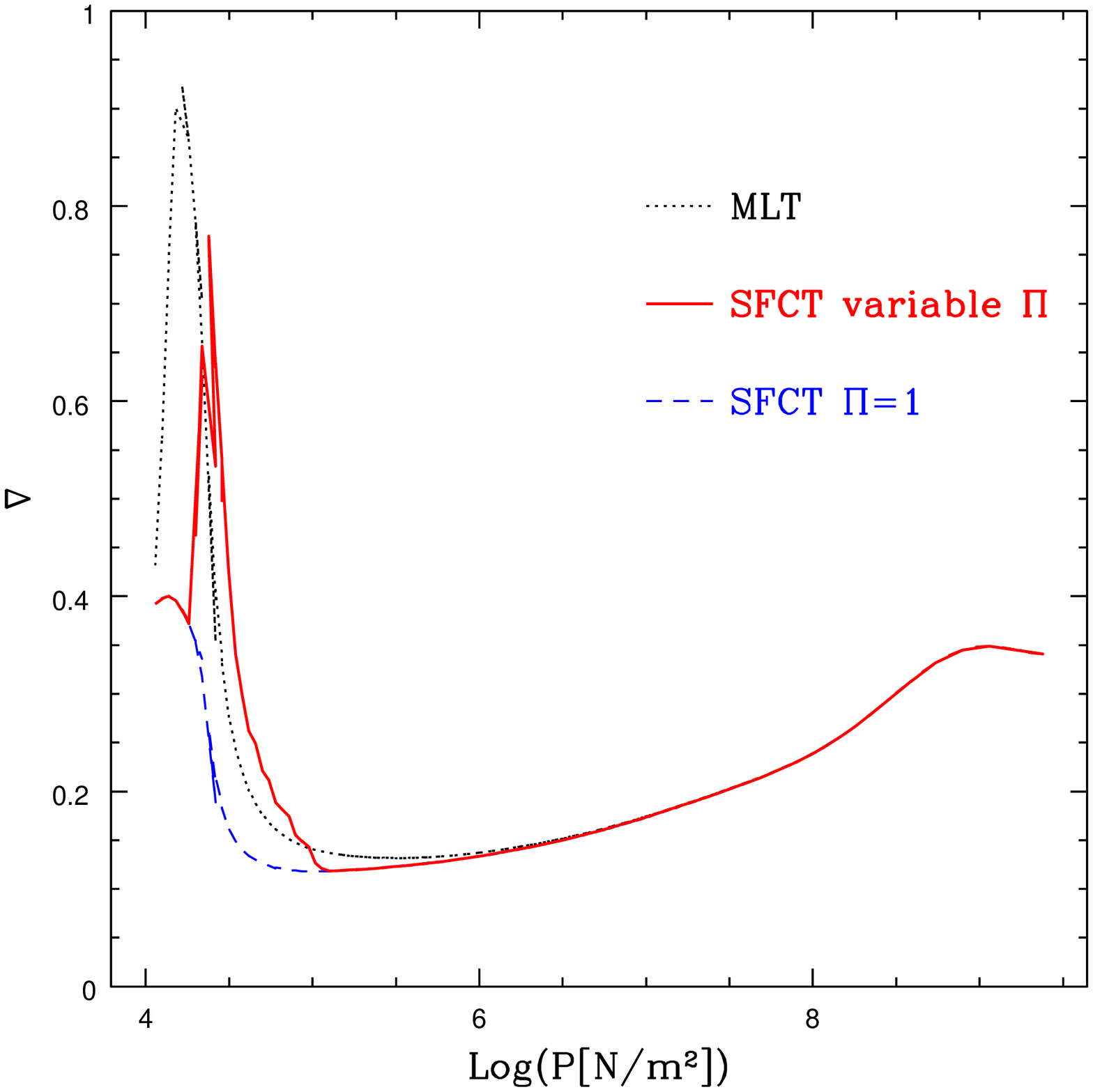}}
\caption{(top panel) The profile of the convective velocity as a function of the pressure across the atmosphere of the $1.0\, M_\odot$ star with initial chemical composition $[X=0.703,\, Y = 0.280,\, Z = 0.017]$,
$log L/L_{\odot}=0$ and age of about 4.6 Gyr,  our best candidate to disposal that should fit the position of the Sun on the HRD. Three profiles are shown; the one derived from the ML theory (black dotted line), the one derived from the SFC theory (red solid line) when the velocity at all layers is let reach the asymptotic regime ($\Pi=1$ everywhere), and finally the third one (blue dashed line) when the velocity in the outer layers can only reach a fraction of the asymptotic value ($\Pi < 1$ in the outer layers).  (bottom panel) The same as in upper panel but for the temperature gradient $\nabla$ of the medium. }
 \label{vel_nabla}
\end{figure}

\section{SFC theory vs. ML theory on the surface boundary conditions}{\label{vis_a_vis}
The  luminosity and effective temperature of a star of mass $M_*$ and chemical composition $[X,Y,Z]$ depend on the energy production (the luminosity) and the energy transport (the effective temperature). The latter, in turn, depends on the combined effect of the radiative and convective transport in the stellar atmosphere, and the very outer layers of this in particular.

What is the behaviour of the SFC theory at this external boundary condition and how to treat it?
The theory developed is fully dynamical, i.e. it includes explicitly the time. Hence careful boundary conditions have to be accommodated to avoid to apply the theory where it loses physical significance, i.e. every time the evolution does not reach the 'asymptotic regime'. In the very external layers,  a convective elements cannot travel  and/or expand upward beyond  the surface of the star. This greatly reduces the  dimension and velocity and lifetime in turn of an element that cames into existence close to the star surface. Therefore it is likely that close to surface, the maximum time allowed to an element is shorter than the time required to reach the asymptotic values of all characteristic physical quantities of the element, the velocity in particular. As a side implication of these considerations, we expect that the resolution of the simulation, i.e. the mass and size zoning, number of mesh points etc.. of the integration technique, should also play a key role in this issues. In other words, we expect a complex interplay between  the mathematical and the numerical technique  employed to simulate the stellar environment embedding the system Eq.\eqref{MySystem}  and  the fundamental physics describing  the intrinsically dynamical nature of the convection theory in use. This is expected also from the physical  fact that the theory works on stellar layers not too large compared to spatial scale over which the  gradients in the main quantities become relevant, but large enough to contain a number of convective elements well represented by statistical indicators (as mean, dispersion etc.). This condition can possibly be missed at the boundary of the star (centre or surface).

The careful analysis of the  outermost layers of stellar atmospheres reveals that while the values and profile of the velocity as a function of the position derived from the ML theory and SFC theory are nearly comparable, those for the ambient temperature gradient $\nabla$  derived from the ML theory greatly differs from  the corresponding ones obtained from the SFC theory. This is shown in Fig.\ref{supplots} for one of  external layers of the $1.0 \, M_\odot$ with the chemical composition $[X=0.703,\, Y = 0.280,\, Z = 0.017]$ on the main sequence. The layer in question is located at the outer edge of the super-adiabatic zone and it is characterized by  $R=6.06736\times 10^8$ m, $T = 7.19449\times 10^3$ K, $P =1.78649\times 10^4 \,\rm Newton\, m^{-2}$, $\rho  = 3.81944\times 10^{-4} \rm \, kg \, m^{-3}$, $\kappa  = 3.25837\times 10^{-1} \, \rm m^2\, kg^{-1}$, $\nabla_{{\rm{ad}}}= 0.345$, $\nabla_{{\rm{rad}}}  = 1.204$, and $\mu = 1.277$.
As in Fig. \ref{soloplots},   a very fine time spacing is adopted. Finally, the meaning of the symbols is the same as in Fig. \ref{soloplots}. Also in this case we compare the SFC theory results with those from the ML theory. Looking at the velocity (left panel), there is coincidence between the SFC theory  and ML theory for the high value past the age of a few $10^3$ s, whereas
for $\nabla$  at the time of a few $10^3$ s, the solutions from the ML theory and the SFC theory show the minimum difference however without reaching coincidence, whereas they strongly deviate both for lower and higher values of the time (similar behaviour is found in other model atmospheres that are not shown here for the sake of brevity).

To lend further support to the above results, we look at the systematic variation of both velocity and ambient temperature gradient across the external convective zone of the model with chemical composition $[X=0.703,\, Y = 0.280,\, Z = 0.017]$ and $\rm{log} \frac{L}{L_{\odot}}=0$ at the age of 4.6 Gyr, our best candidate to disposal to fit the position of the Sun on the HRD.    The results are presented in Fig.\ref{vel_nabla} and are compared to those of the ML theory.   The red solid lines show the asymptotic velocity and companion $\nabla$ derived from the straight  application of the SFC theory. In the case of the ML theory (the black dots), the concept of an asymptotic regime for the velocity and $\nabla$ in turn does not apply because given the physical condition of the medium there is only one, time-independent value for both the velocity and $\nabla$.  Velocities and temperature gradients in the deep regions of the convective zone predicted by the SFC theory and ML theory are nearly coincident whereas toward the surface they tend to greatly differ.  The temperature gradients of the SFC theory is much lower than the one of the ML theory.
Too low a value for the ambient $\nabla$  would immediately imply a smaller radius and a higher effective temperature in turn  (the luminosity being mainly driven by the internal  physical conditions is hardly affected by what happens in the atmosphere).  The immediate consequence is that the final position on the HRD of the evolutionary track is too blue to be able to match the Sun. Similar results are found also for models of star of different mass and evolutionary stage.

From this analysis, we learn that not all layers of a convective regions, those near the stellar surface in particular, can reach the  asymptotic regime for the convective velocity (and also all other relevant quantities). To clarify this important issue, we proceed as follows.

We start  calculating the time at which the  solution of the quintic equation satisfies all the conditions of Eq.\eqref{Conds01} (e.g., with $\Pi=0.95$). This usually occurs when  the time is about $10^5 - 10^6$ s and sometime less in the outermost layers. We name this time "numerical time $t_{\rm{asy}}$". It is not a physical time but a numerical-method dependent variable: different $\Pi$ fixed arbitrarily  give different $t_{\rm{asy}}$. Then we argue that in a convective region, owing to continuous upward/downward motion of the fluid elements, the effect of any variation/perturbation of the physical quantities will soon or later propagate throughout the  convective region at a speed whose maximum value is the sound speed $v_s= \sqrt{\Gamma_1 P/\rho}$ (with the usual meaning of all the symbols)\footnote{ It is worth noting here that since we are dealing we the most external layers in which the ionization of light elements (H,He, C, N, O, etc...) takes place, the expression for $\Gamma_1$ to use is the one containing the effect of ionization as well as radiation pressure. See Appendix B or \citet[][]{1968pss..book.....C} for details on the expression for $\Gamma_1$ we have used.}.
Suppose now that the whole convective region has a width $\Delta r_{\rm{cnv}}$. At each layer of the convective zone, we may calculate the sound velocity $v_s$ and associate a temporal time-scale $t_{\rm{cnv}}$, i.e. the time-scale a convective element  would require to expand its size to the whole convective region.
At each layer we have that  the convective element expansion rate ${\dot \xi _e} \to v_s $ and the maximum size $\Delta  {\xi}_e$ satisfies the condition $ \Delta {\xi}_e = \Delta r_{\rm{cnv}}/2$ so that we  define $t_{cnv}$ as
\begin{equation}\label{tconv}
{t_{{\text{cnv}}}} \sim \frac{{\Delta {{  \xi }_e}}}{{{{\dot \xi }_e}}} = \frac{{\Delta {{  \xi }_e}}}{{{v_s}}} = \frac{\Delta r_{\rm{cnv}}}{2 v_s} \, .
\end{equation}
At any layer the asymptotic regime cannot be reached  if the two time-scales  are in the ratio
\begin{equation}\label{asy_conv}
	{ \frac{t_{\rm{asy}}}{t_{\rm{cnv}}}} = \Pi < 1.
\end{equation}
This condition fixes also the maximum fraction of the local velocity with respect to its asymptotic value reached in  each layer. The percentage $\Pi$ varies with the position. Going deeper into the star the sound velocity increases,  $t_{\rm{cnv}}$ decreases, and the  condition (\ref{asy_conv}) is always violated, i.e. the asymptotic velocity is reached anyhow. In these regions  the ratio $\Pi$ always  reaches the maximum value $\Pi=1$.

From a technical point of view, the extent of the convective region  $\Delta r_{\rm{cnv}}$ is not known a priori and therefore an iterative procedure must be adopted starting from a reasonable guess. Basing on the calculations of many model atmosphere,   $\Delta r_{\rm{cnv}}$, we expressed starting guess value for the convergence as $\Delta r_{conv} \simeq h_p \times N$, where $h_p$ is the pressure scale height of the outermost layer and   $N$ the typical number of mesh points describing an atmosphere when $\rm{\log} P$ is the independent variable. This finding greatly facilitates the task of choosing the initial guess  for $\Delta r_{\rm{cnv}}$. One or two iterations of the atmosphere are sufficient to refine $\Delta r_{\rm{cnv}}$ to the desired value. Our model atmospheres are calculated with $N\simeq 150$ mesh points. Using different codes with different resolving algorithm and numerical precision,  different values of $N$ can be found.

The procedure we have described acts as numerical scheme for the boundary conditions on the velocity profile in the outermost regions of a star. To illustrate the point, in Fig.\ref{profile} we show the   profile of $\Pi$ throughout the atmosphere of the zero age main sequence model of the $1\, M_\odot$ star. It is worth noting that the asymptotic value of the velocity can be reached everywhere except  in the outermost layers of the star for $\log P < 6\, \rm N \, m^{-2}$. As far as we can tell, this behaviour is the same in stellar models of the same mass but in different evolutionary stages and in models in the same evolutionary stage but different mass.
\begin{figure}
\centering
\resizebox{\hsize}{!}{\includegraphics{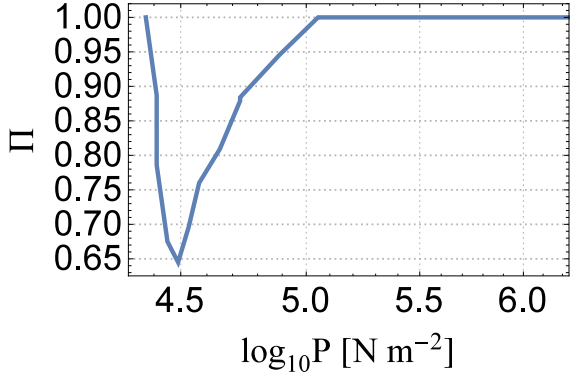}}
\caption{The profile of $\Pi$ across the atmosphere of the MS model of the $1.0 M_\odot$ star with chemical composition $[X=0.703,\, Y = 0.280,\, Z = 0.017]$. Note the fall  of $\Pi$to the minimum value  followed by the rapid increase to $\Pi=1$  at increasing pressure.}
 \label{profile}
\end{figure}
The number $N$ of mesh points  in the atmosphere is not a free parameter, but it is fixed by the mathematical technique and accuracy of the integration procedure (in our case $N\simeq 150$) and therefore it cannot be changed without changing these latter. Other stellar codes should have different values of $N$.   However, limited to the following discussion  we will take advantage of relations \ref{tconv} and \ref{asy_conv}  to assess the model response to variations of $\Delta r_{\rm{con}}$, $\Pi$, velocity $v$ of the convective elements  and finally $\nabla$ by simply varying $N$. At each layer, keeping the sound velocity $v_s$ constant (the physical quantities $P$, $\rho$ etc. are assigned)  lower $N$s   would imply smaller $\Delta r_{\rm{cnv}}$,  higher values of $\Pi$ and velocity in turn,    too low $\nabla$'s in outermost layers, and eventually too blue evolutionary tracks in the HRD  with respect to those from the ML theory. The opposite is the case for  higher values of $N$.
\begin{figure}
\centering
\resizebox{\hsize}{!}{\includegraphics{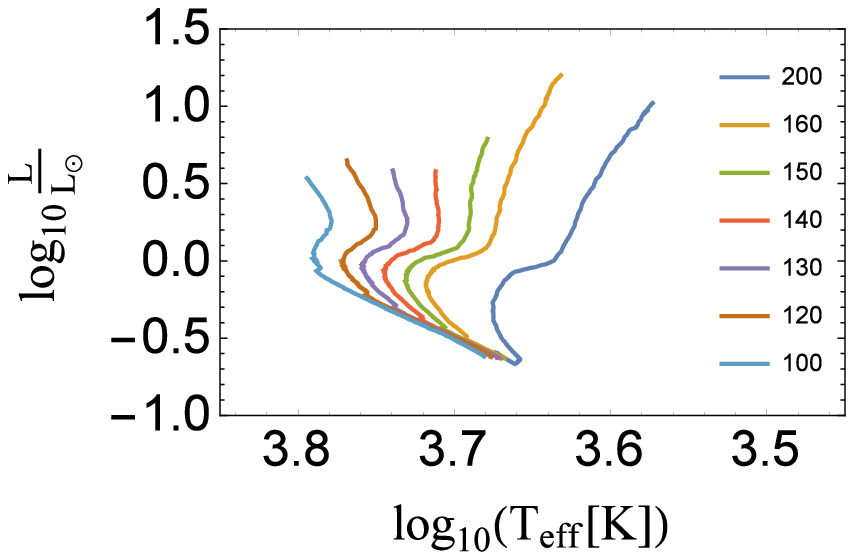}}
\resizebox{\hsize}{!}{\includegraphics{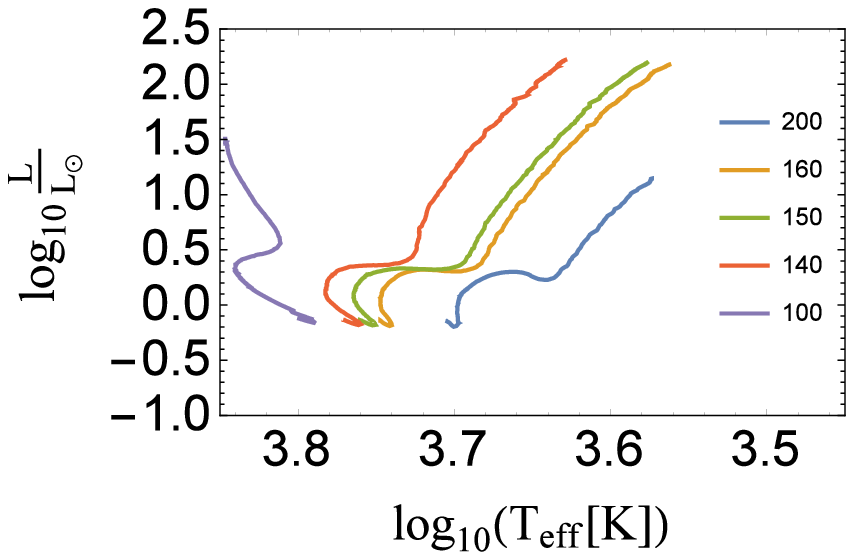}}
\caption{Evolutionary sequences of the 0.8$M_\odot$ (Top Panel) and 1.0 $M_\odot$ (Right Panel) stars with chemical composition $[X=0.703,\, Y = 0.280,\, Z = 0.017]$ at artificially varying the number $N$ of mesh points in the atmosphere as indicated. Only the case with $N=150$  and the right value of $\Delta r_{cnv}$  shows a track  in agreement with current results in literature. The quintic is solved with time resolution $\Delta e=0.01$. As explained in great detail in the text, low values of $N$ correspond to small size of the convective region,  high values of the velocity, smaller radii and hence higher effective temperatures. The opposite is the case of large values of $N$. Therefore only the correct $\Delta r_{cnv}$, profile $\Pi$,  and velocities in the region of strong super-adiabaticity yield stellar models and evolutionary tracks in agreement with real HRDs. In other words, only the correct application of the SFC theory yields results able to reproduce the observations.}
 \label{test_seq}
\end{figure}
The results of these numerical experiments  are shown in Fig. \ref{test_seq} for a  test evolutionary sequences for the 0.8 and 1$\, M_\odot$ with chemical composition $[X=0.703,\, Y = 0.280,\, Z = 0.017]$, that are calculated forcing a variation of the size $\Delta r_{\rm{cnv}}$  by varying the number $N$ as indicated.  In conclusion, the correct physical description of the outermost layers of the external convective region of a star is crucial to  calculate  stellar models able to reproduce the position of real stars on the HRDs.

\section{Results}\label{Results00}
As already recalled in the previous sections, both the classical ML theory and the new SFC theory find their best application in the convective regions of the the outer layers of a star where the super-adiabatic convection occurs. Therefore, first we  investigate the physical structure of model atmospheres that are calculated both with the standard ML theory and  the new SFC theory.

The numerical code for the atmosphere models has been extracted from the classical G\"ottingen code developed by  \citet{1964ZA.....59..215H} and  used and implemented by the Padova group for more than four decades.
Over the years, this code has been developed to include semi-convection \citep[e.g.,][]{1970Ap&SS...8..478C}, ballistic convective overshoot from the central core  \citep{1981A&A...102...25B}, envelope overshoot \citep[][]{1991A&A...244...95A}, turbulent diffusive mixing and overshoot \citep[][]{1996A&A...313..145D,1996A&A...313..159D,1999A&A...342..131S}, and finally the many revisions of the input physics  and improvements described in \citet[][]{1994MmSAI..65..689B}, \citet[][]{1995A&A...301..381B}, \citet[][]{2001AJ....121.1013B}, \citet[][]{2003AJ....125..770B}, \citet[][]{2008A&A...484..815B}.
The version used here is the one by \citet{1994A&AS..106..275B}    in which we have replaced the ML theory with the SFC theory.
The value of $\Lambda_m$ adopted for the ML theory is taken from \citet{2008A&A...484..815B} and provides calibrated models matching  the properties of the the Sun on the HRD. The adopted value is $\Lambda_m=1.68$. The structure of the atmosphere models  is according to the equations and physical input described in Sections \ref{Outerlayers}, \ref{ML theory}, and \ref{NewTheory}.
\begin{figure*}
\centering
{\includegraphics[width=15.0truecm, height=15.0truecm]{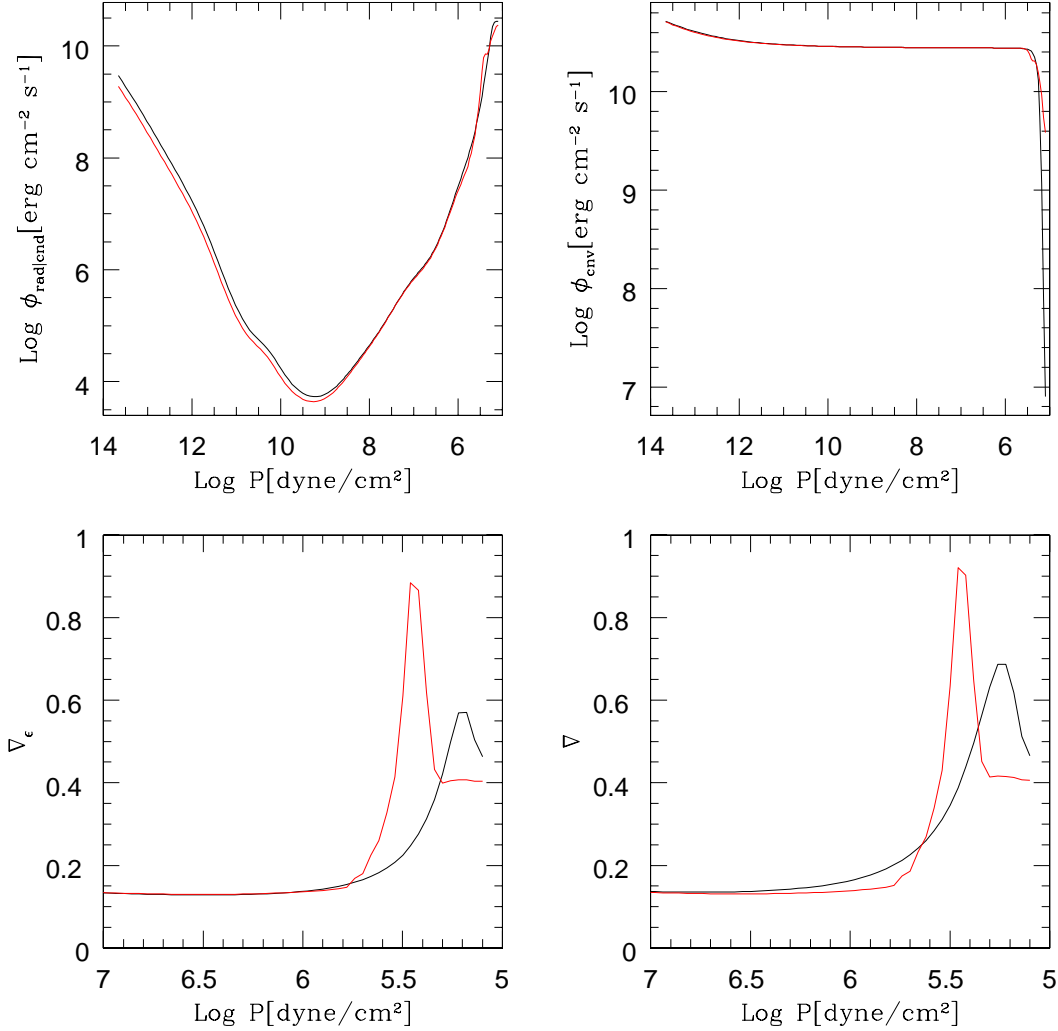}}
 \caption{Structure of the outer layers of the Sun. Solar fluxes and temperature gradients profiles for the internal convective stratification of the star. The upper panels show  the radiative flux ${\varphi _{{\rm{rad|cnd}}}}$ (left)  and the convective flux ${\varphi_{{\rm{cnv}}}}$ (right). The bottom panels display the element gradient $\nabla_e$ (left) and the ambient gradient $\nabla$ (right). The red lines refer to SFC  theory whereas the black lines to the ML theory.}
 \label{Fig_sun_now}
\end{figure*}
\subsection{The outer layers of Sun-like stars: atmosphere models}\label{TheSun}
We take the evolutionary track of $\rm 1\,M_\odot$ with chemical composition [X=0.71, Y=0.27, Z=0.02] calculated by \citet[][]{2008A&A...484..815B} and isolate the model that fairly matches the position of the Sun on the HRD, i.e.
$ \log L/L_ \odot$=0 and  $\log T_{\rm{eff}}$=3.762.  The atmosphere is shown in Fig. \ref{Fig_sun_now}. In each panel we show the results for the standard ML theory (using  ${\Lambda} = 1.68$ and the SFC theory. We display the radiative flux ${\varphi _{{\rm{rad|cnd}}}}$ (top left panel) and the convective flux ${\varphi _{{\rm{conv}}}}$ (top right panel), the  logarithmic temperature gradient of the element $\nabla_e$ (bottom left panel) and of the medium $\nabla$, (bottom right panel). The colour code indicates the underlying theory of convection, black  for the ML theory and red  for the new theory.
It is soon evident that while the profiles of the fluxes are virtually identical with the two theories, the gradients $\nabla_e$ and $\nabla$ are much different, a result already visible in Fig. \ref{vel_nabla}. In both cases, the  extent of  convective  zones is similar. By construction the position on the HRD is the same.
For the sake of illustration we show in Fig. \ref{Fig_star_tip} the case of a $2.0 M_\odot$ star with same chemical composition and in an advanced stage along the RGB, the luminosity is $\log L/L_{\odot} = 2.598$ and the effective temperature $\log T_{\rm eff}=3.593$. The situation is much similar to the previous one.
\begin{figure*}
\centering
{\includegraphics[width=15.0truecm, height=15.0truecm]{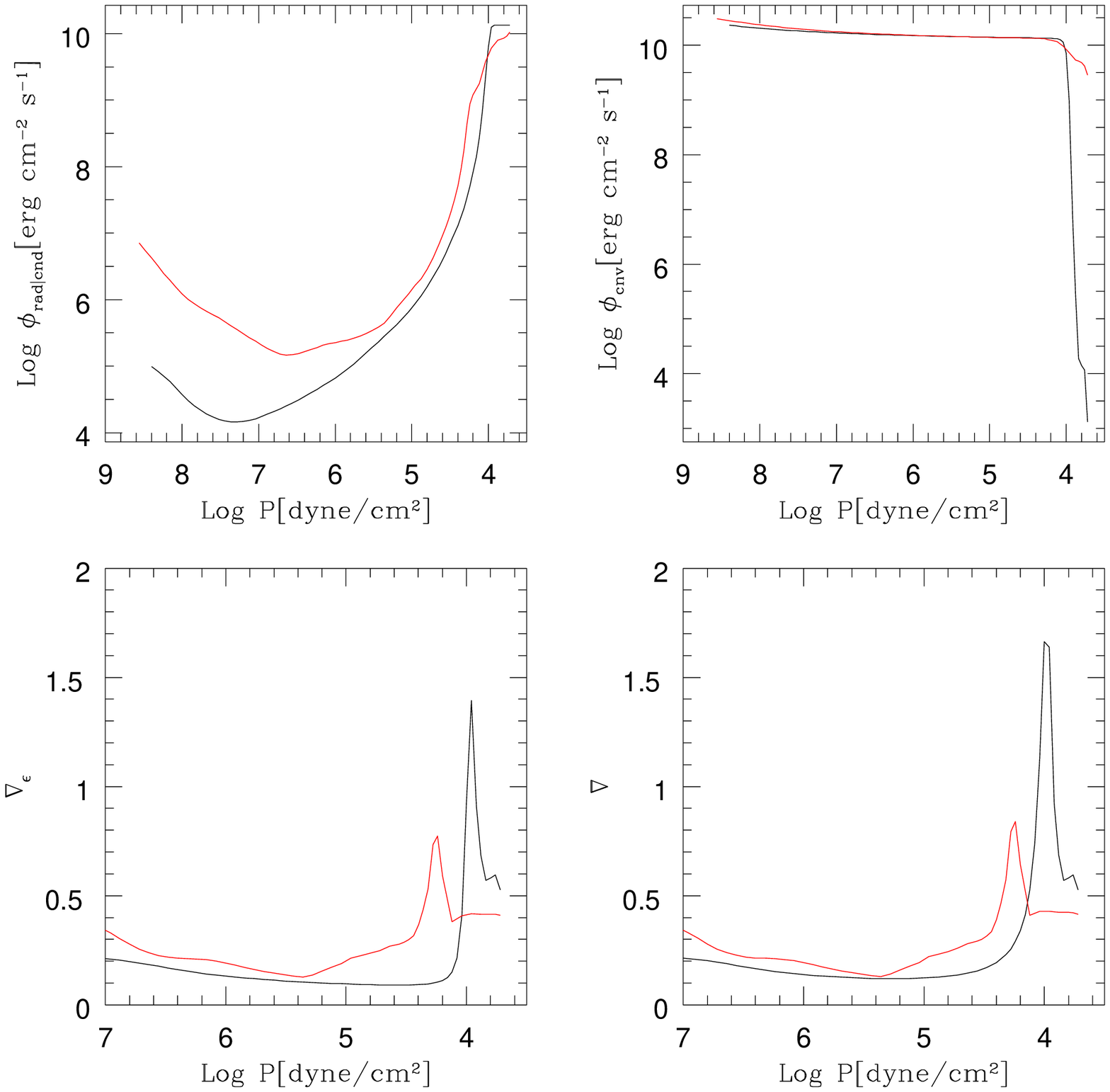}}
 \caption{The same as in  Fig \ref{Fig_sun_now} but for the 2$M_\odot$ in a late stage along the RGB,
 $\log L/L_\odot$=2.598 and  $\log T_{\rm{eff}}$=3.593.}
 \label{Fig_star_tip}
\end{figure*}
The  new SFC theory yields the same path on the HRD as the calibrated MLT, however the greatest merit of former is that no ML parameter or calibration is required. The properties of convection are fully determined by the physics of the layer in which convection is at work. By taking the outer envelope of the model whose luminosity and effective temperature are those of the Sun and looking at the stratification of the main variables (temperature, density pressure, radiative and convective fluxes, velocity  and associated dimensions of convective elements, temperature gradient  in presence of convection), it is soon evident that the ML theory is indeed a particular case in the more general solutions predicted by the SFC theory. As we go deeper into the atmospheres, the solutions for the ML theory and SFC theory tend to diverge. This is expected and it simply reflects the fact that these external solutions are not constrained to match the inner solution at the transition layer (typically $M/M_* \simeq 0.97$, where $M$ and $M_*$ are the mass at the layer $r$ and total mass, respectively). Small differences among the two solutions tend to amplify as we go deeper inside. This is more evident in case of the $2.0 M_\odot$ star along the RGB.
This can be fixed only by considering complete stellar models. Hence a few preliminary exploratory stellar models will be presented below.

\begin{figure*}
\centering
\resizebox{\hsize}{!}{\includegraphics{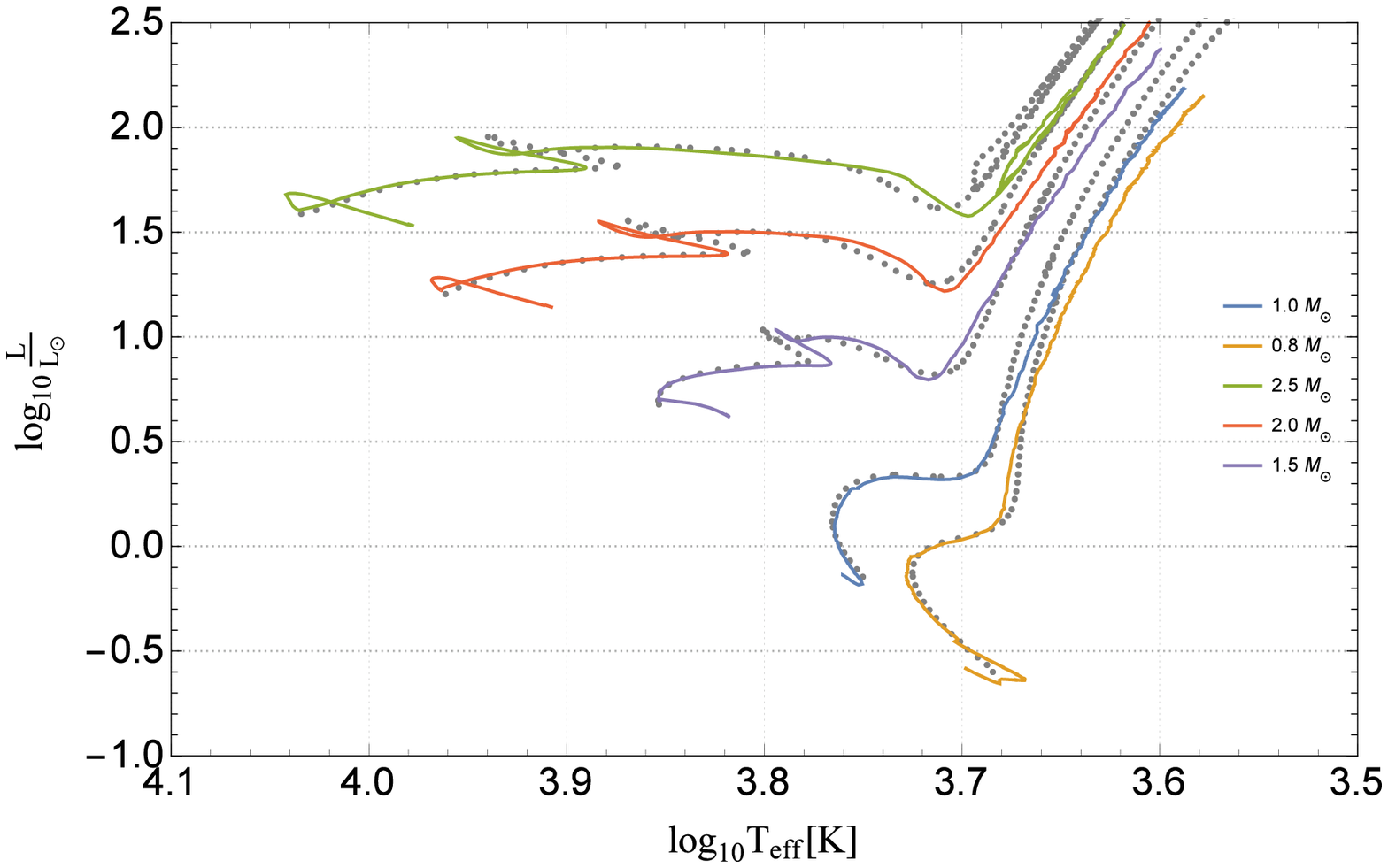}}
\caption{ The HRD of the 0.8, 1.0, 1.5, 2.0, and 2.5 $M_\odot$ stars with initial chemical composition $[X=0.703, Y = 0.280, Z = 0.017]$ calculated from the main sequence to advanced evolutionary stages using both the classical ML theory (the crossed lines) and the SFC theory (dotted lines of different colors). The 0.8, 1.0, 1.5 and 2.0 $M_\odot$ models are carried to a late stage of the RGB before core He-ignition (He-Flash), whereas the 2.5 $M_\odot$ is evolved up to very advanced stages of central He-burning ($Y_c \simeq 0.1$). The stellar models are calculated with the Padova code and input physics used by \citet{1994A&AS..106..275B} and \citet{2008A&A...484..815B}, see also the text for more details. The models are meant to prove  doubt that that the SFC theory with no ML  parameter is perfectly equivalent to the classical ML theory with calibrated  ML  parameter ($\Lambda_m=1.68$ in our case). }
 \label{Good_HRD}
\end{figure*}

\subsection{Preliminary, complete stellar models with the SFC theory}
We have calculated a few test evolutionary sequences of complete stellar models for different initial mass and fixed chemical composition. The stellar models are followed from the main sequence stage up to the end of the RGB  or core helium exhaustion, as appropriate to the initial mass of the star.

Several important remarks  are  mandatory here before presenting the stellar models under consideration. First the SFC theory we have described  is specifically designed to deal only with convection in the outer layers of the stars: it cannot be applied to deal with physical situations in which convective overshooting either from central cores and/or convective intermediate shells is taking place. However, we would like to mention that
the formalism developed by \citet{2014MNRAS.445.3592P} derives the acceleration acquired by convective elements under the action of the buoyancy force in presence of the inertia of the displaced fluid  and gravity. Therefore, it is best suited, with the necessary modification, to derive the motion of the convective elements beyond the formal limit set by the Schwarzschild condition, the penetration of these into the surrounding radiative regions, the dissipation of their kinetic energy and finally the redistribution of energy and physical properties of the layer interested by their motion, i.e. to describe convective overshooting. We are currently working on extending   SFC theory to the convective overshooting \citep[][in preparation]{Pasettoetal2016}.
Therefore, in order to calculate new stellar models with the SFC theory, we must use one of the prescriptions  for convective overshooting from the core currently in literature. We adopt here the ballistic model of convective overshooting developed by \citet{1981A&A...102...25B} end  since adopted by the Padova group. It is not the best solution but it is sufficient to obtain significant exploratory results.

The code considered in this models is the same from which we have taken all the routines to calculate the model atmospheres. All the input physics, i.e. opacities (radiative conductive and molecular), nuclear reaction rates, equation of state, and the prescription for convective overshooting from the core are as described by \citet[][]{2008A&A...484..815B}, to whom the reader should refer for all details.
In particular, it is worth recalling that
the treatment of core overshooting relies on \citet{1981A&A...102...25B} that stands on the ML theory (to derive the velocity of convective elements) and makes use of the ML  parameter $\Lambda_c=0.5$ for all masses $M_* \geq 1.5 \, M_\odot$, $\Lambda_c =0$ for stars with mass  $M_* \leq 1 \, M_\odot$, and finally $\Lambda_c = M_*/M_\odot -1.0$ for stars in the interval  $1.0 < M_*/M_\odot \leq 1.5$. Overshooting from the bottom of the convective envelope along the RGB follows from \citet{1991A&A...244...95A} with $\Lambda_e=0.25$.  Therefore, the interiors are calculated according to the classical prescription, whereas the outer layers are treated according to the SFC theory. This is   an  intermediate step towards the correct approach in which convective overshoot in the internal regions is treated in the framework of SFC theory.

We note that an obvious drawback of  using the \citet[][]{1994A&AS..106..275B} code is that the input physics is somewhat out of date with respect to more recent versions of the same code, eg.
\citet{2008MmSAI..79..738N}, \citet{2008A&A...484..815B},   \citet{2009A&A...508..355B},  and finally the very recent revision of the whole code by  \citet{2012MNRAS.427..127B}, \citet{2014MNRAS.444.2525C} and \citet{2014MNRAS.445.4287T}. The choice of the \citet[][]{1994A&AS..106..275B}
code is motivated by the large body of stellar models calculated with this and worldwide used. In any case, this satisfactorily permits the comparison of stellar models with the same code, input physics and both ML theory and SFC theory. Work is under way  to calculate new grids of stellar models with SFC theory using an independent code with very modern input physics i.e. the Garching code named GARSTEC by  \citet{2008Ap&SS.316...99W}.

For the purposes of this exploratory investigation we present here five evolutionary sequences for stars of initial mass  0.8, 1.0, 1.5, 2.0, and 2.5  $M_\odot$ and  chemical composition [$X=0.703, Y = 0.280, Z = 0.017$] calculated from the main sequence to advanced evolutionary stages using both the classical ML theory ($\Lambda_m=1.68$ in our case) and the SFC theory. The 0.8, 1.0, and 1.5 $M_\odot$ sequences  are indicative of the old stars in Globular Clusters and very old Open Clusters, whereas the 2.0 and 2.5 $M_\odot$ sequences correspond to intermediate age Globular and Open clusters. The 2.0 $M_\odot$ is the last  low mass star of the adopted chemical composition undergoing core He-Flash \citep{2008A&A...484..815B}.  The HR Diagram  is shown in Fig. \ref{Good_HRD}, where the grey dots indicate the sequences with the ML theory and the dotted lines of different colours show those with the SFC theory. The 0.8,   1.0, 1.5, and 2.0 $M_\odot$ models are carried to a late stage  of the RGB before core He-ignition (He-Flash), whereas the 2.5 $M_\odot$ is evolved up to very advanced stages of central He-burning, $Y_c \simeq 0.1$ (no He-Flash has occurred).
The corresponding models with the classical ML theory (dotted paths) are taken from \citet{2008A&A...484..815B}.

In general the two sets of models are in close  agreement. However looking at the results in some detail, the new tracks tends to have a slightly different inclination of the RGB. The SFC tracks are nearly identical to those of the ML theory at the bottom  and progressively becomes redder towards the  top, i.e. the RGBs of the low mass stars are less steep than those of the classical MLT models. Looking at the case of the $1\, M_\odot$ star, the MLT model are calculated with $\Lambda_m=1.68$ upon calibration on the Sun and kept constant up to the end of the RGB and afterwards. The models with the SFC theory do not require the mixing length parameter but fully agree with  the MLT ones during the core H-Burning phase but by the time they reach the RGB tip they would be in better agreement with MLT models with a smaller values of the mixing length parameter. The required decrease of $\Lambda_m$ is difficult to quantify at the is stage of model calculations. However, it agrees with the analysis  made by
with  \citet{2015A&A...573A..89M} of
3D radiative hydrodynamic  simulations of convection in the envelopes of late-type stars in terms of the 1D classical ML theory. Using different calibrators and mapping the results a s function of gravity, effective and effective temperature \citet{2015A&A...573A..89M} find that  at given gravity the ML parameter increases with decreasing effective temperature, the opposite at given effective temperature and decreasing gravity. There are also additional dependencies  on metallicity and  stellar mass that we leave aside here. Looking at the case of the Sun, passing from the main sequence to a late stage on the RGB, the ML is found to decrease by as much as about 10 percent. Applying this to stellar models, a less steep RGB would result as shown by our model calculations with the SFC theory.
Owing to the complexity of the new SFC theory with respect to the classical ML theory, the results are very promising.   These preliminary model calculations show that that the SFC theory with no ML  parameter is  equivalent to the classical ML theory with calibrated  ML. More work is necessary to establish  a quantitative correspondence between the two theories of convection.

\section{Conclusions and future work}\label{Conclusion}
We have presented here the first results of the integration of stellar atmospheres and exploratory full stellar models to which  the new convection theory developed by \citet{2014MNRAS.445.3592P} has been applied. To this aim, a mathematical and computational algorithm and a companion code have been  developed to integrate the system of equations governing the convective and radiative fluxes, the temperature gradients of the medium and elements and finally, the typical velocity and dimensions  of the radial and expansion/contraction motion of convective elements. In parallel we have also calculated the same quantities with the standard ML theory in which the ML  parameter has been previously calibrated.  All the results obtained with ML theory are recovered with the new theory but no scale parameters are adopted. We claim that the new theory is able to capture the essence of the convection in stellar interiors without a fine-tuned parameter inserted by hand.

The main achievement of the theory presented in this paper is not only to prove that satisfactory results can be achieved, as was already done by the ML theory, but more importantly to clarify that our understanding of the stellar structure is correct and fully determined by the underlying physics.
Each star ``knows its own convection'': i.e, where it is located, how much it extends and how much energy it is able to carry away.  This is the meaning and the power of the self-consistent results we have just presented.
Finally, the theoretical picture we have developed  has a  predictive power that merely descriptive analyses of numerical simulations still miss. In other words, successful numerical experiments of laboratory hydrodynamics with millions of degrees of freedom  do  not   imply a complete understanding of the phenomenon under investigation. An emblematic example of this  is offered by the impressive simulations by  \citet{2015arXiv150300342A}.  However, even in this case the closure of the basic equations involved in their  hydrodynamic simulations is not possible.  This has been instead achieved by the much simpler and straightforward  formulation of the same problem by  \citet{2014MNRAS.445.3592P}. Based on these preliminary results we are confident that this is the right path to follow.
However, before moving toward more complicated physical situations such as convective overshoot and semi-convection and extending our  theory to deal with these  phenomena \citep{Pasettoetal2016}, it is necessary to check the overall consistency of the new theory by calculating  stellar models over all possible evolutionary phases according to the mass of the star, to extend the calculations to wider ranges of initial masses (namely to massive stars where mass loss by stellar winds is important all over their evolutionary history and  the very low mass ones  where convection is becoming more and more important), and finally to consider other initial chemical compositions.  Work is in progress to this aim \citep{Pasettoetal2016}.

Finally, a numerical  code for the solution of the polynomial \eqref{Quintic02} or \eqref{Quintic01} both in $ \xi_e$ and $\left| {v}\right|$ is available upon request to the first author.

\section{Acknowledgments}

C.C. warmly acknowledges the support from the Department of Physics and Astronomy of the Padova University. E.C. thanks the hospitality of the INAF-Astronomical Observatory of Padova.

\begin{small}
\bibliographystyle{mn2e}                    
\bibliography{Pase_Conv_II_Biblio}                 

\end{small}

\begin{appendix}

\section{The external layers: basic equations and input physics}

\subsection{The Photosphere}\label{photosphere}
Given the total mass $M_*$ and the chemical composition [X, Y, Z] (following traditional notation, X represent the H concentration, Y the He and Z the remanent elements so that $X+Y+Z=1$ identically holds), adopted the spherical symmetry, and a system of polar coordinates ${\bf{x}} = \left\{ {r,\theta ,\phi } \right\}$ centred on the barycentre of a star, the boundary conditions at the surface of a star are
\begin{equation}
r=r_*  \qquad    T_{\rm{sup}}=T_{\rm{ph}}   \qquad \rho_{\rm{sup}}=\rho_{\rm{ph}}.
\end{equation}

\noindent
Determining $T_{\rm{ph}}$ and  $\rho_{\rm{ph}}$ (or $P_{\rm{ph}}$) is not simple, requiring instead a detailed treatment of the most external layers of a star. The photosphere, which corresponds to  the surface of a star, is defined as the most external layer from which the radiation coming from inside is eventually radiated away and above which the Local Thermodynamic Equilibrium can no longer be applied. It is the last layer at which the radiation is nearly identical to that of a black body at the temperature $T$. This layer is also used to define the effective temperature $T_{\rm{eff}}$ implicitly as:
\begin{equation}
 L=4\pi~r_*^{2}\sigma T_{\rm{eff}}^{4}.
\end{equation}
The photosphere is also the layer at which the matter is no longer transparent to radiation as it occurs far away from the star. We now make use of the concept of optical depth at the photon frequency $\nu$, $\kappa_\nu $, Eddington approximation for the radiative transfer equation and grey body. The momentum flux equation for a photon fluid reads
\begin{equation}\label{Atmo01}
	\frac{{\partial \varphi }}{{\partial t}} + \left\langle {\nabla ,{\bm{{{\bm{P}}_\nu }}}_{\rm{rad}}} \right\rangle  =  - \rho \kappa_\nu \varphi,
\end{equation}
where $\varphi$ is the radiative flux of photons and ${\bm{{{\bm{P}}_\nu }}}_{\rm{rad}}$ is the pressure radiation tensor for photons of frequency $\nu$. Eddington introduced the function ${K_\nu }$ which is widely used in literature and related to the outward component of the monochromatic pressure tensor ${P_{{\rm{rad}}{\rm{,}}\nu }}$ of frequency $\nu$ as ${P_{{\rm{rad}}{\rm{,}}\nu }}=4 \pi{K_\nu } $ together with two functions: $H_\nu$ (the Eddington flux) and $F_\nu$ related by ${P_{{\rm{rad}}{\rm{,}}\nu }} = 4\pi {H_\nu } = \pi {F_\nu }$\footnote{These functions can be proved to be simply statistical moments of the intensity weighted by $\cos \theta$ and $\cos^2 \theta$ related to the radiative flux and the radiative pressure in any direction $\theta$ pointing outside the star to the observer.}. Making use of the grey-body approximation (i.e., independence of the specific frequency $\nu$) and this notation, Eq.\eqref{Atmo01} reads
\begin{equation}\label{Edd01}
	\frac{{dK}}{{d\tau }} = \frac{F}{4}.
\end{equation}
Because we assumed radiative equilibrium, we can integrate the above equation to obtain the mean intensity $  I = \frac{1}{{4\pi }}\oint {Id\Omega }$ over the solid angle ${d\Omega }$ as
\begin{equation}
	  I = \frac{3}{4}F\left( {\tau  + Q} \right),
\end{equation}
where the Eddington condition $  I = 3K$ has been employed on Eq.\eqref{Edd01} and Q is the integration constant. For a linear intensity relation, ${I_\nu } = a\mu  + b$ with $\mu  = \cos \theta$ cosine director outward the star pointing to us, we can fit the Sun limb darkening with $\frac{a}{b} \cong \frac{3}{2}$ and determine the constant $Q$ (normalized to the Sun) as $Q \cong \frac{2}{3}$. Finally, if we assume that in the stellar layer considered the mean-free-path of a photon is much smaller than the characteristics scale length where the temperature changes, ${\lambda _\nu } = \frac{1}{{\rho \kappa }} \ll {h_T}$, the diffusion-approximation applies ($  I \cong \frac{a}{{4\pi }}{T^4}$) and with the definition of effective temperature above, we obtain:
\begin{equation}
{T^4} = \frac{3}{4}\left( {\tau  + \frac{2}{3}} \right)T_{{\rm{eff}}}^4.
\end{equation}
Therefore, the photosphere is the layer whereby  $\tau={2/3}$. A more rigorous solution by \cite{1960ratr.book.....C} \citep[see also][pg.62]{1982stat.book.....M} sees the factor $\frac{2}{3}$ in the previous equation substituted by $q\left( \tau  \right) \in \left] {0.577,0.710} \right[$. This relation  yields the dependence of the temperature on the optical depth in the region $\tau=0$ (${\rho}=0$) to $\tau={2/3}$ in the so-called grey atmosphere approximation (i.e. $\kappa$ independent from the radiation frequency).

Finally, the pressure at the photosphere is given by the hydrostatic equilibrium condition as a function of  $\tau$,
\begin{equation}
 P_{\rm{ph}}={\frac{G m}{r^2}} \int_0^{\tau_{\rm{ph}}}{\frac{1}{\kappa}} d\tau,
\label{c18_11}
\end{equation}
where  $P_{\rm{ph}}=0$ at $\tau=0$  and radiation pressure is neglected and $G$ the gravitational constant.
The opacity $\kappa$ is a function of the position, $\kappa=\kappa(r)$, and therefore the state variables $P$, $T$, $\rho$, and chemical composition $\mu$. However, to a first approximation $\kappa$ can be considered constant and taken equal to the value at the photosphere. It follows from this that
\begin{equation}
 P_{\rm{ph}}={\frac{2}{3}}{\frac{1}{\kappa_{\rm{ph}}}}{\frac{G m}{r^2}}
\label{c18_12}
\end{equation}
Note that the effect of radiation pressure the can be absorbed by recasting in the previous equation the gravity as $g_{\rm{eff}} = g - \frac{\kappa \sigma T_{\rm{eff}}^4}{c}$ with $g = \frac{{GM}}{r^2}$.
The  relationships for $T_{\rm{ph}}$ and  $P_{\rm{ph}}$ define the natural boundary conditions for the  system of equations  describing a  stellar structure.
To conclude, temperature, pressure and density in the regions above the photosphere are  expressed as functions of the optical depth $\tau$, whereas below the photosphere to be determined they require the complete set of stellar structure equations.

\subsection{The atmosphere}\label{atmosphere}
In absence of rotation, magnetic fields and in hydrostatic equilibrium, the structure of a star is defined by the following equations:

\begin{itemize}
	\item The mass conservation
			\begin{equation}
			\frac{{dM_r}}{{dr}} = 4\pi {r^2}\rho,
			\label{mass_con}
			\end{equation}
where $M_r$ is the mass inside the sphere of radius $r$.
	\item The gravitational potential ${\Phi _{\bf{g}}}$ satisfying the Poisson relation
			\begin{equation}
			 {\frac{\partial \Phi_{\bf{g}}}{\partial r}} = 4\pi g \rho.
            \label{poten}
			\end{equation}
	\item The condition of mechanical equilibrium for a fluid at rest (Euler equation):
			\begin{equation}
				\frac{{d P}}{{d r}} = - \rho \frac{G M_r}{r^2},
			\label{hydro}
            \end{equation}
			with $G$ is the gravitational constant.
  \item The equation for the energy conservation
			\begin{equation}
			\frac{{d L_r}}{{d r }} = 4 \pi r^2 \rho(\varepsilon_N  - {\varepsilon _\nu } + {\varepsilon _g}),
			\label{lum_eq}
			\end{equation}
		where $\varepsilon_N$,  ${\varepsilon _\nu }$ and  ${\varepsilon _g}$ are the nuclear, neutrino losses, and 				 gravitational sources, respectively, and $L_r$ is the luminosity from the sphere of radius $r$.
	\item Finally, the equation for energy transfer, which can be expressed as follows
			\begin{equation}
			 \frac{d\ln T}{d\ln P} = \nabla,
			\label{ene_trans}
			\end{equation}
\end{itemize}
where $\nabla$ depends on the dominating physical mechanism for energy transport: $\nabla_{{\rm{rad/cnd}}}$ for radiation plus conduction, $\nabla_{\rm{cnv}}$ for convection (typically in stellar atmospheres), and simply $\nabla_{{\rm{ad}}}$ in presence of adiabatic convection (typically in deep stellar interiors).

In the atmosphere, first it is more convenient to use the pressure $P$ instead of the radius $r$  as the independent variable, and second  all the three energy sources $\epsilon_n$, $\epsilon_g$, and $\epsilon_\nu$ can be assumed to be zero so that the luminosity is a constant $L(M)=L=\rm{const.}$. Consequently Eq. (\ref{lum_eq}) is no longer needed and
Eqs. (\ref{mass_con}), (\ref{hydro}), (\ref{ene_trans}) read:
\begin{equation}\label{sysApp1}
\left\{ \begin{array}{rcl}
  \frac{d \ln M_r}{d \ln P} &=& - \frac{4\pi r^4 P}{G M_r^2} \\
  \frac{d \ln r}{d \ln P} &=& -\frac{r P}{G \rho M_r} \\
   \frac{d \ln T }{d \ln P} &=& \nabla,
\end{array} \right.
\end{equation}
with $L(P) = L = \rm{const.}$ respectively.

These equations must be complemented by the EoS in the atmosphere, $P=P(\rho, T, \mu)$ with $\mu$ the molecular weight of the chemical mixture (inclusive of ionization), the opacity $\kappa(\rho, T , \mu)$,  and the expressions for $\nabla$ that depend on the transport mechanism at work.
If $\nabla_{{\rm{rad}}} < \nabla_{{\rm{ad}}}$, the energy flows by radiative transport and the temperature gradient of the medium is
\begin{equation}\label{radGradB}
\nabla = \nabla_{{\rm{rad}}} = \frac {3}{16\pi a c G } \frac{\kappa L P }{M T ^4},
\end{equation}
with $a$ density-radiation constant and $c$ speed of light.
If $\nabla_{{\rm{rad}}} \geq \nabla_{{\rm{ad}}}$ convection sets in. The energy flux is carried by radiation and convection.
Let us indicate with $\varphi$,  $\varphi_{\rm{cnv}}$ and  $\varphi_{\rm{rad|cnd}}$ the total energy flux, the convective flux, and the radiative  plus conductive (if needed) energy flux lumped together\footnote{Conduction has an important role in the degenerate cores of red giants and advanced stages of intermediate-mass and massive stars, and dominates  in the isothermal cores of white dwarfs  and neutron stars. The conductive flux can be expressed by the same relation for the radiative flux provided the opacity is suitably redefined. In the external layers of a normal star conduction  in practice has no role and the above notation is superfluous. However in view of the future extension of the SFC theory to internal convection and/overshooting, we keep also here this more general notation.}. Among the three fluxes the obvious equation $\varphi = \varphi_{{\rm{cnv}}} + \varphi_{\rm{rad|cnd}}$ applies.  In this region four temperature gradients are at work:  the gradient of convective elements $\nabla_e$, the gradient of the medium in presence of convection $\nabla_{\rm{cnv}}$, the adiabatic gradient $\nabla_{{\rm{ad}}}$,  and a fictitious gradient still named $\nabla_{{\rm{rad}}}$ as if all the energy flux were carried by radiation.  While the flux carried by radiation is easily known, the flux carried by convection requires a suitable theory to specify $\nabla$ and $\nabla_e$.
The above system of Eqs.(\ref{sysApp1})  together with those describing the convective transport represent the ambient of  the stellar atmosphere in which  super-adiabatic convection is at work either according to the  ML theory or the new SFC theory that will be shortly summarized in  Appendix \ref{MLtheory_SFCtheory}.

To complete the physical description of the stellar medium, we need to  present here  a few important thermodynamic quantities that are used to derive temperature gradients and the convective flux in presence of ionization and radiation pressure.

\subsection{Ionization and Thermodynamics of an ionizing gas}\label{ion_therm}
To proceed with the calculation of  $\nabla_e$ and $\nabla$  required by the systems of Eqs. \eqref{MLtheoryeq1} for ML theory or \eqref{MySystem} for the SFC theory we need $\nabla_{\rm{ad}}$ and $c_{\rm{P}}$  for a gas made of a number elemental species in various degrees of ionization and in presence of radiation pressure. Despite several formulations of this equation exist in literature \citep[e.g.,][]{1962ZA.....54..114B,1964ZA.....59..215H,1994sse..book.....K,2012sse..book.....K,1968pss..book.....C}, we present here the basic equations adopted in this paper. They are taken from \citet{1962ZA.....54..114B} however  adapted to our notation and strictly limited to those used in our code.

\subsubsection{Ionization}\label{ioniza}
Consider a mixture of atoms of type $i=1,...,N$, each of which with  $n_{e,i}$ electrons and  $n_{e,i}+1$ stages of ionization indicated by $r=0,...,n_{e,i}$ (we neglect here the case of atoms in a give stage of ionization but different state of excitation), the fraction of atoms of type $i$ in the $i^{th}$ stage of ionization (i.e. that have lost $i$ electrons) is $x_i^r$. The total fraction $y_i^j$ of atoms of type $i$ which are in ionization stages higher than $i^{th}$  is $y_i^r = \sum_{s=r+1}^{n_{e,i}} x_i^s$.
Let us indicate the relative number of atoms of type $i$ as $\nu_i=n_i/n$ with $n$ total atoms of $N$ types. The total fraction $f_e$ of free electrons is then:
\begin{equation}\label{FracNe}
 f_e= \sum_{i=1}^N \nu_i \sum_{r=0}^{n_{e,i}} r x_i^r = \sum_{i=1}^N \sum_{r=0}^{n_{e,i} -1} \nu_i y_i^r  \,.
\end{equation}
We introduce the function
\begin{equation}
K_i^r \equiv \frac{u_{r+1}}{u} \frac{2}{P_{\rm{gas}}} \frac{(2\pi m_e)^{3/2}(k_B T)^{5/2}}{h^3}\, e^{- \frac{\chi_i^r}{k_B T}},
\label{kap_ir}
\end{equation}
for $ r=0,1, .....Z_{i-1}$, where $\chi_i^r$ is the $r^{th}$ ionization potential of atom $i$, $u$ the statistical weight of the state $r$ $k_B$ the Boltzmann constant and $h$ the Plank constant. Then, to derive the degree of ionization we need to solve the system of $ \sum_{i=1}^N (n_{e,i} -1) $ Saha's equations together with the $N$ equations $ \sum _{r=0}^{n_{e,i}} x_i^r=1$ for the $\sum_{i-1}^N n_{e,i}$ quantities $X_i^r$:
\begin{equation}\label{sysApp}
\left\{ \begin{array}{rcl}
	\frac {x_i^{r+1} } { x_i^r}  \frac {f_e} {f_e+1} &=& K_i^r \\
	 \sum _{r=0}^{n_{e,i}} x_i^r &=& 1,
\end{array} \right.
\end{equation}
that can be solved  numerically. In most cases, however, the ionization potentials $\chi_i^r$ differ sufficiently from each other so that only one ionization is taking place at any time. Therefore for a given $i=k$ only $x_k^s$ and $x_k^{s+1}$ are different from zero. The condition ($\sum _{r=0}^{n_{e,i}} x_i^r = 1$ can then be approximated by $x_i^s + x_i^{s+1} =1$ and accordingly $y_k^r = 0$ for $r > s$, $y_k^s = x_k^{s+1}$, and $y_k^r =1$ for $r<s$, and the Saha's equation becomes a quadratic expression:
 \begin{equation}
  \frac{y_k^s}{1-y_k^s} \frac{A_k^s + \nu_k Y_k^s}{1 + A_k^s + \nu_k y_k^s} = K_k^s,
 \end{equation}
with the aid of the auxiliary quantity
 \begin{equation}\label{AFracNe}
 A_k^s \equiv \sum_{i\neq k}^N \sum_{r=0}^{n_{e,i} -1} \nu_i y_i^r   + \nu_k \sum_{r \neq s} ^{n_{e,i} - 1} y_k^r,
 \end{equation}
that has fully algebraic solution.

\subsubsection{Thermodynamics}\label{thermo}
The derivation of $\nabla_{\rm{ad}}$ and $c_{\rm{P}}$ accounting for the effect of ionization  is  as follow. In Eq.\ref{FracNe} and \ref{AFracNe} we absorb the indexes over the atoms and ionization, i.e. we write simply $f_e=\sum_{i=1}^N \nu_i y_i $ and $A_k \equiv \sum_{i\neq k}^N  \nu_i y_i $. Then, for any stellar layer the specifi heat $c_p$ is
 \begin{equation}\label{Cpionized}
{c_P} = \frac{\Re }{{{\mu _0}}}\left( {\frac{5}{2} + \frac{{4(1 - \beta )(4 + \beta )}}{{{\beta ^2}}}} \right)\left( {1 + {f_e}} \right) + \sum\limits_i {\frac{{{\nu _i}}}{{{G_i}}}F_i^2},
 \end{equation}
where the auxiliary functions $F_i$ and $G_i$ are
\begin{equation}\label{sysApp2}
\begin{array}{rcl}
{F_i} &\equiv& \frac{5}{2} + \frac{{4\left( {1 - \beta } \right)}}{\beta } + \frac{{{\chi _i}}}{{{k_B}T}}\\
{G_i} &\equiv& \frac{1}{{{y_i}\left( {1 - {y_i}} \right)}} + \frac{{{\nu _i}}}{{{f_e}\left( {1 + {f_e}} \right),}}
\end{array}
\end{equation}
where $1 - \beta  = \frac{{a{T^4}}}{{3P}}$, $\mu  = \frac{{{\mu _0}}}{{1 + {f_e}}}$, $\delta  =  - {\left( {\frac{{\partial \ln \rho }}{{\partial \ln T}}} \right)_P}$, and $\alpha  =  - {\left( {\frac{{\partial \ln \rho }}{{\partial \ln P}}} \right)_T}$ have been used. Finally, under the same hypotheses, the adiabatic gradient is
\begin{equation}\label{Adiabo}
 \nabla_{{\rm{ad}}} = \frac{\left(1 + \frac{(1-\beta)(4+\beta)}{\beta^2}\right)(1+f_e) +\frac{1}{\beta} \sum_i \frac{\nu_i}{G_i} F_i } {\left(\frac{5}{2} + \frac{4(1-\beta)(4+\beta)}{\beta^2}\right)(1+f_e) + \sum_i \frac{\nu_i}{G_i} F_i^2 }.	
\end{equation}
where $\nabla_{{\rm{ad}}}$  is 0.4 for a perfect neutral gas with no radiation, tends to 0.25 for a fully ionized gas in presence of radiation, and may further decrease to about 0.12 in presence of  ionization as in the case of external layers. Finally, the generalized adiabatic exponent $\Gamma_1$ (that is needed to calculate the sound velocity) is
\begin{equation}\label{Gamma1}
\begin{array}{l}
{\Gamma _1} = \left( {2\beta (3\beta (\beta  + 8) - 32)({f_e} + 1){G_i} - 4{\beta ^3}{F_i}^2{\nu _i}} \right) \times \\
\left( {{\nu _i}\left( {\beta \left( {2{F_i} + 3} \right) - 8} \right)} \right.\left( {\left( {8 - \beta \left( {2{F_i} + 3} \right)} \right) - 8\beta } \right) + \\
{\left. { + 6\beta \left( {7\beta  - 8} \right)\left( {{f_e} + 1} \right){G_i} + \beta {\nu _i}\left( {\beta \left( {4{F_i} + 39} \right) - 64} \right)} \right)^{ - 1}}.
\end{array}
\end{equation}
All the   model atmospheres used in this study are calculated including radiation pressure and ionization of light elements and the effect of these on all thermodynamical quantities in use.   For more details the reader should refer to the original sources
\citep{1962ZA.....54..114B,1964ZA.....59..215H,1994sse..book.....K,2012sse..book.....K,1968pss..book.....C}.

\section{The ML and SFC theories of convection}\label{MLtheory_SFCtheory}
The above equation of stellar structure in the atmosphere require a suitable theory of convection. In this appendix first we summarize the version of the classical ML theory we have adopted and then we shortly review the new SFC theory of \cite{2014MNRAS.445.3592P}
In what follows we will omit all demonstrations and intermediate passages to focus the attention on the basic assumptions and main results.

\subsection{Mixing-length theory: a summary} \label{ML theory}
The equations for the energy flux transport  of the ML theory are available from literature in several forms (but equivalent in content). They are:
\begin{equation}\label{MLtheoryeq1}
\left\{ \begin{array}{rcl}
{\varphi _{{\rm{rad|cnd}}}} &=& \frac{{4ac}}{3}\frac{{{T^4}}}{{\kappa {h_p}\rho
}}\nabla \\
{\varphi _{{\rm{rad|cnd}}}} + {\varphi _{{\rm{cnv}}}} &=&
\frac{{4ac}}{3}\frac{{{T^4}}}{{\kappa {h_p}\rho }}{\nabla _{{\rm{rad}}}}\\
{{  v}^2} &=& g\delta \left( {\nabla  - {\nabla _e}} \right)\frac{{l
_m^2}}{{8{h_p}}}\\
{\varphi _{{\rm{cnv}}}} &=& \rho {c_P}T\sqrt {g\delta } \frac{{l _m^2}}{{4\sqrt
2 }}h_p^{ - 3/2}{\left( {\nabla  - {\nabla _e}} \right)^{3/2}}\\
\frac{{{\nabla _e} - {\nabla _{{\rm{ad}}}}}}{{\nabla  - {\nabla _e}}} &=&
\frac{{6ac{T^3}}}{{\kappa {\rho ^2}{c_P}{l_m}{  v}}},
\end{array} \right.
\end{equation}
where $h_p$ is the scale height of the pressure stratification of the star, $v$ the average velocity of the convective element, and  all other symbols have their usual meaning. In particular we recall that  $l_m$  is the mean dimension and mean free path of the convective elements before dissolving and releasing  their energy excess to the surrounding medium. It is customarily expressed as $l_m = \Lambda_m h_p$, where $\Lambda_m$  is the mixing-length parameter. The  derivation and solution  of this system of equations can be found in  any classical textbook of stellar structure \citep[e.g.][]{1994sse..book.....K,1964ZA.....59..215H,2012sse..book.....K}. In literature, there are several versions of the ML theory \citep[see][and references]{2014MNRAS.445.3592P} but in this paper we prefer to follow the one presented by \citet{1964ZA.....59..215H} and ever since adopted by the Padova group in their stellar evolution code to calculate the structure of the most external layers of a star \citep[see][and references]{2008A&A...484..815B}.

The set of Eqs. (\ref{MLtheoryeq1})  can be lumped together in a dimensionless equation among the three gradients $\nabla$ (of the medium in presence of convection), $\nabla_{{\rm{rad}}}$, and $\nabla_{{\rm{ad}}}$. Introducing the quantity
\begin{equation}\label{effic}
\begin{array}{rcl}
V  &\equiv& \frac{{3ac{T^3}}}{{{c_P}{\rho ^2}\kappa l_m^2}}\sqrt {\frac{{8{h_p}}}{{g\delta }}}  \\
W &\equiv& {\nabla _{{\rm{rad}}}} - {\nabla _{{\rm{ad}}}},
\end{array}
\end{equation}
we may  derive from system Eq. \eqref{MLtheoryeq1}  the dimensionless equation
\begin{equation}\label{equa_conv}
{\left( {\xi  - V } \right)^3} + \frac{8}{9}V \left( {{\xi ^2} - {V ^2} - W} \right) = 0,
\end{equation}
where $\xi$ is the positive root of ${\xi ^2} = \nabla  - {\nabla _{{\rm{ad}}}} + {V ^2}$. The solution of Eq. \eqref{equa_conv} is algebraic. Writing the Eq. \ref{equa_conv} in standard form
\begin{equation}
	{\xi ^3} - \frac{{17{U^3}}}{9} - \frac{{19U}}{9}{\xi ^2} + 3{U^2}\xi  - \frac{{8UW}}{9} = 0,
\end{equation}
and using the  Tschirnhaus transformation, $\xi  = \eta  + \frac{{19U}}{9}$, we get
\begin{equation}
{\eta ^3} + \frac{{368}}{{243}}{U^2}\eta  - \frac{{9344}}{{19683}}{U^3} - \frac{8}{9}UW = 0.
\end{equation}
Writing this equation in the compact form  ${\eta ^3} + py + q = 0$, the associated discriminant is $\Delta  =  - 4{p^3} - 27{q^2} < 0$. Thereore, we expect the solutions of Eq.\ref{equa_conv} to have  only one real root given by:
\begin{equation}
	\eta  = \sqrt[3]{{\frac{1}{2}\left( { - q + \sqrt {{q^2} + \frac{4}{{27}}{p^3}} } \right)}} + \sqrt[3]{{\frac{1}{2}\left( { - q - \sqrt {{q^2} + \frac{4}{{27}}{p^3}} } \right)}}
\end{equation}
with $p = \frac{{368}}{{243}}{V^2}$ and $q =  - \frac{{9344}}{{19683}}{V^3} - \frac{8}{9}VW$. Inverting the Tschirnhaus transformation we obtain the final solution.
Once $\nabla$ is known, one may derive $\nabla_e$ from the relation $ \nabla_e - \nabla_{{\rm{ad}}} = 2V \sqrt{\nabla - \nabla_e}$, so that the four gradients and  the fluxes $\varphi_{{\rm{rad}}}$ and $\varphi_{{\rm{cnv}}}$ are determined,  and the whole problem is solved.
Despite this apparently simplicity, in the literature there are several different expressions for the coefficient of the cubic equation \cite[e.g.,]{2012sse..book.....K, 1968pss..book.....C, 2009pfer.book.....M} or even for the  equation system Eq.\eqref{MLtheoryeq1}. We will adopt the solution obtained from the above equations.

The drawback of the ML theory is the ML parameter that cannot be determined in the framework of the ML theory itself. Vice versa the theory proposed by \citet{2014MNRAS.445.3592P} tries to describe the motion of convective elements taking into account that  in addition to the upward /downward motion due to the buoyancy force they also expand/contract while moving so that they are subjected to other effects. By doing this, new equations are found which together with those based on the energy conservation lead to a self-consistent description of the motions of convective elements without introducing arbitrary free parameters. As expected, the physics of the medium itself determines all the properties of  convection at each unstable layer of a star.

\begin{figure}
\includegraphics[width=\columnwidth]{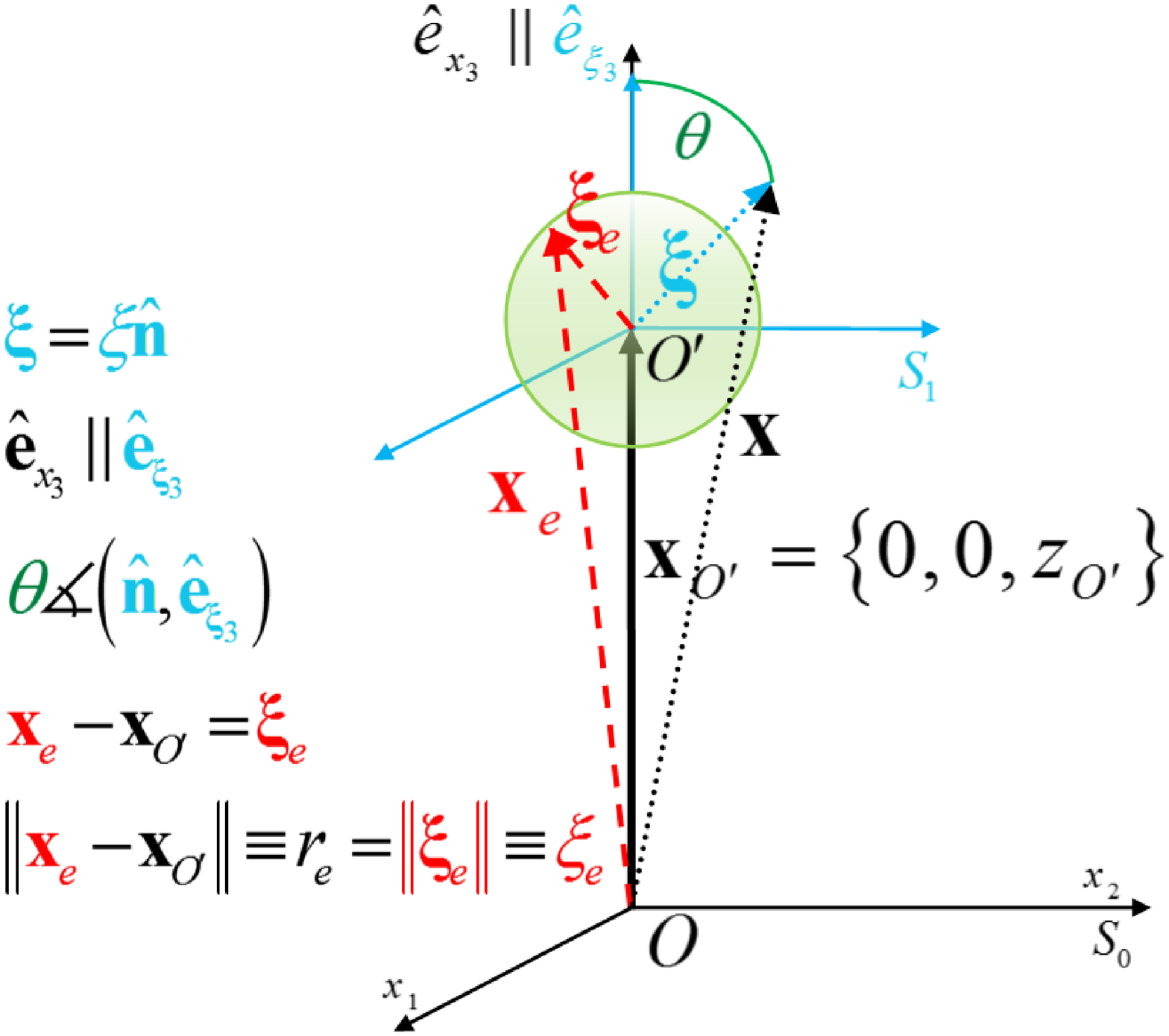}
\caption{Schematic representation of a convective element seen in the inertial frame $S_O$ involved and in the co-moving frame $S_1$. The element is represented as spherical body for simplicity. The center of the sphere indicated as  $O'$ corresponds also to the position of the element in $S_0$.  The generic dimension of the convective element as seen in $S_1$ is indicated by $\xi_e$. }
\label{Fig1}
\end{figure}

\subsection{The  new theory of stellar convection}\label{NewTheory}
The key idea of the new theory of stellar convection by \citet{2014MNRAS.445.3592P} is extremely simple. Let us consider a rising  convective element. In an ideal star, because of the spherical symmetry, the  motion occurs along the radial direction, while at the   same time the element increases its dimension.
The opposite happens for an element sinking into the medium: we have radial motion and shrinkage.  Because upward (downward) motion and expansion (shrinkage) of the element are intimately related (indeed the element rises because it expands and sinks because it shrinks). We remind the reader that in the classical ML theory only the radial motion is explicitly considered whereas expansion and shrinkage although implicitly present are not taken into account. We emphasize that the presence of the ML parameter simply mirrors the incomplete description of the motion of convective elements that is limited to the radial direction. Therefore the natural trail to follow to develop an alternative scale-free theory of convection is to look at the  expansion/contraction, the radial motion being physically connected.
The goal can be easily achieved if instead of using the natural reference frame $S_0$ centred on a star's center (inertial system), we make use of a frame of reference $S_1$ centered  on and co-moving with the generic convective element (non-inertial system). In $S_1$, the element is at rest with respect to the surrounding medium while it expands / contracts into it. The two reference frames are schematically shown in Fig. \ref{Fig1}.  In this case  the motion of a generic element can be described by the integral of the Navier-Stokes equations, i.e. the Bernoulli equation,  in which neglecting magnetic fields and viscous terms (typical of high-Reynold-Number fluids in which viscous terms are small compared to inertia terms), the velocity potential approximation can be adopted.
In the following, we sketch the SFC theory by \citet{2014MNRAS.445.3592P} highlighting  the main hypotheses, fundamental equations  and key results. No demonstration is given.

\subsubsection{Formulation of the problem and basic equation}
As already said, the stellar medium is a perfect fluid with a suitable EoS function of time $t$ and position $\bm x$ as viewed in the inertial system $S_0$  of Fig.\ref{Fig1}. A perfect fluid is intrinsically unstable and turbulent, therefore the higher the Reynolds number the better the above approximation holds good.
Furthermore, on macroscopic scales the stellar interiors  are represented by a   perfect fluid in mechanical and thermodynamical equilibrium;  all other forces (viscous, centrifugal in presence of rotation by rotation, and electromagnetic in presence of magnetic fields) but gravity and pressure gradient are neglected; on large scales the fluid is incompressible and irrotational \footnote{The concept of a large distance scale for incompressibility and irrotationality is defined here from a heuristic point of view: this length scale should be large enough to contain a significant number of convective elements so that a statistical formulation is possible when describing the mean convective flux of energy (see below), but small enough so that the distance travelled by the convective element is short compared to the typical distance over which significant gradients in temperature, density, pressure etc. can develop (i.e. those gradients are locally small).}, Finally, the concept of potential flow can be exploited here: the velocity field can be derived from the gradient of suitable potential \citep[see][chap. 1]{1959flme.book.....L}.
 In $S_1$, combining the  Euler's and mass conservation equation, we can obtain the Bernoulli equation for not inertial reference frames as \cite[]{2012A&A...542A..17P}:
\begin{equation}\label{Eq003}
	\frac{{\partial {\Phi _{{{\bm{v}}_0}}}}}{{\partial t}} + \frac{P}{\rho } + \frac{{{{\left| {{{\bm{v}}_0}} \right|}^2}}}{2} + {\Phi _{\bm{g}}} = f\left( t \right) -\left\langle {{\bf{A}},{\bf{\xi }}} \right\rangle
\end{equation}	
where $\Phi_{{\bm{v}}_0}$  is the velocity potential generating the  fluid velocity
 ${\bm{v}}_0$  and ${\Phi _{\bm{g}}}$ the gravitational potential. This relation describes the stellar plasma in which convection is at work.

The main target of any theory of stellar convection is to find solutions of Eq. (\ref{Eq003}) linking  the physical quantities characterizing the stellar interiors  such as pressure, density, temperature, velocities etc. and the mechanics governing the motion  of the convective elements as functions of  the fundamental temperature gradients with respect to pressure introduced above, i.e. the radiative gradient ${\nabla _{{\rm{rad}}}}$, the adiabatic gradient ${\nabla _{{\rm{ad}}}}$, the local gradient of the star $\nabla $, the convective element gradient ${\nabla _e}$ and the molecular weight gradient ${\nabla _\mu } $. The task is very difficult. The problem however can be tackled in a rather simple fashion making use of the velocity potential.\\

\subsubsection{Velocity potential in an accelerated frame $S_1$}\label{VElpot}
Let us now introduce the reference frame $S_{1}: (O',{\bm{\xi }})$  co-moving with and centred on the center of the generic element.
From the geometry shown in Fig. \ref{Fig1}, the radius of a generic convective element of spherical shape is  indicated as $\left|\bm{x_e - x_{O'}}\right| =\bm{r_e}$ in $S_0$ and $\left|\bm{x_e - x_{O'}}\right| =\bm{\xi_e}$ in $S_1$.
\citet{2014MNRAS.445.3592P} have demonstrated that the total potential flow outside the surface of the moving and expanding/contracting elements in $S_1$ is given by
\begin{equation}\label{Eq004}
\Phi^\prime =  - \left\langle {{\bm{v}},{\bm{\xi }}} \right\rangle \left( {1 + \frac{1}{2}\frac{{\xi _e^3}}{{{{\left| {\bm{\xi }} \right|}^3}}}} \right) - \frac{{{{\dot \xi }_e}\xi _e^2}}{{\left| {\bm{\xi }} \right|}},	
\end{equation}
so that the corresponding velocity in ${S_1}$ can be written as
\begin{equation}
\label{Eq005}
{{{\bm{v^\prime}}}_0} = {\left. {\frac{3}{2}\left( {\left\langle {{\bm{v}},{\bm{\hat n}}} \right\rangle {\bm{\hat n}} - {\bm{v}}} \right) + {{\dot \xi }_e}{\bm{\hat n}}} \right|_{\left| {\bm{\xi }} \right| = {\xi _e}}},
\end{equation}
with meaning the symbols as in Fig.\ref{Fig1}. The above expression is evaluated  at the surface of the convective. It is also easy to show that this equation yields correct results at the surface of the element once written in spherical coordinates with $\theta$ the angle between the unitary vectors $\bm \hat e_z$ and $\bm\hat \xi$.
Finally, the time derivative  of Eq. (\ref{Eq003})  is
\begin{equation}\label{Eq006}
{\left. {\frac{{\partial \Phi '}}{{\partial t}}} \right|_{\left| {\bm{\xi }} \right|={\xi _e}}} =
- \frac{3}{2}{\xi _e}\left\langle {{\bm{A}},{\bm{\hat n}}} \right\rangle  - \frac{5}{2}{\dot \xi _e}\left\langle {{\bm{v}},{\bm{\hat n}}} \right\rangle  - {\ddot \xi _e}{\xi _e} - 2\dot \xi _e^2	
\end{equation}
where the relative acceleration of the two reference frames is indicated with ${\bm{A}}$.
The inclusion of Eq.\eqref{Eq004}, \eqref{Eq005} and \eqref{Eq006} in Eq.\eqref{Eq003} will lead to the general relation
\begin{eqnarray}\label{Eq010}
\frac{{{v^2}}}{2}\left( {\frac{9}{4}{{\sin }^2}\theta  - 1} \right) &-& v{{\dot \xi }_e}\frac{5}{2}\cos \theta + \left( {\frac{P}{\rho } + {\Phi _{\bm{g}}}} \right) = \nonumber\\
+  A{\xi _e}\left( {\frac{3}{2}\cos \theta  - \cos \phi } \right) &+& {{\ddot \xi }_e}{\xi _e} + \frac{3}{2}\dot \xi _e^2,
\end{eqnarray}
where $A = \left| {\bm{A}} \right|$ is the norm of the acceleration,  $\phi $ the angle between the direction of motion of the fluid as seen from ${S_1}$ and the acceleration direction,  and  $\theta $ the angle between the radius $\xi$ in $S_1$ and the velocity $\bm{v}$. It is interesting  to note that if we consider approximatively equal the pressure above and below the convective element, we can use the previous equation to obtain an relation for the motion of the barycentre of a non-expanding convective element. At an arbitrary point of this rigid-body approximation we get:
\begin{equation}\label{EqIIIA}
	\begin{array}{rcl}
		\frac{{{{  v}^2}}}{2}\left( {\frac{9}{4}{{\sin }^2}\theta  - 1} \right) &=& \frac{{A{{  \xi }_e}}}{2}\cos \phi  \\
		- \frac{{{{  v}^2}}}{2} &=&  \pm \frac{{A{{  \xi }_e}}}{2} \\
		{{  v}^2} &=&  \mp A{{  \xi }_e}, \\
	\end{array}
\end{equation}
which is one of the equations  that we want to integrate. A different derivation of this equation will be given in Section \ref{appendAccel}  below.
Eq.\eqref{Eq010} is the  version in  spherical-coordinates  of a  general theorem \citep[see][Sec. 4.1]{2014MNRAS.445.3592P} whose applicability is large but of little practical usefulness because of its  complexity.  Nevertheless, it is the cornerstone of the new theory.

\subsubsection{The motion-expansion/contraction rate relationship}\label{Secref2.3}
In order to obtain equations analytically treatable, \citet[][]{2014MNRAS.445.3592P} limited their analysis to the linear regime. To this aim they needed  a parameter the value of which remains  small enough to secure the linearization of the basic equations.  If we limit ourselves to \textit{subsonic  stellar convection}, it is assumed that the upward/downward velocity of a convective element, ${\bm{v}}$, will be much smaller that its expansion rate $\left| {\frac{{d{\bm{\dot \xi }}_e}}{{dt}}} \right| \equiv \left| {{\bm{\dot \xi }}_e}\right|$, i.e.
$$\left| \bf{v}\right| <<  \left| {{\bf{\dot \xi_e }}} \right|.$$
This seems to be a reasonable assumption for the majority of the situations we are examining because asymptotically in time the expansion rate of the convective element will tend to the local sound velocity. This allows us to develop  a linear theory based on the small parameter
$\varepsilon  \equiv \frac{ {\left| {\bf{v}  } \right|}}{{\left| {{\bf{\dot \xi_e }}} \right|}} \ll 1$.
In this limit case, Eq.(\ref{Eq010}) becomes
\begin{equation}\label{Eq018}
	{{{\ddot \xi }_e}{\xi _e} + \frac{3}{2}\dot \xi _e^2 + \frac{{A{\xi _e}}}{2} = 0},
\end{equation}
which rules the temporal evolution of the expansion rate of a  convective element. The straight solution of this equation is difficult but feasible. We refer to the Appendix of \citet{2014MNRAS.445.3592P} for all mathematical details.
The asymptotic solution for  $\tau  = \frac{t}{{{t_0}}} \to \infty$  is of interest here and it is given as a function of the dimensionless size of a generic convective element, $\chi  \equiv \frac{\xi }{{{\xi _0}}}$, by:
\begin{equation}\label{Eq031}
\chi \left( \tau  \right) = \frac{1}{{{4}}}{\tau ^2} + \frac{{\sqrt \pi  \Gamma \left( { {\raise0.5ex\hbox{$\scriptstyle 7$}
\kern-0.1em/\kern-0.15em
\lower0.25ex\hbox{$\scriptstyle 8$}}} \right)}}{{\Gamma \left( {{\raise0.5ex\hbox{$\scriptstyle 3$}
\kern-0.1em/\kern-0.15em
\lower0.25ex\hbox{$\scriptstyle 8$}}} \right)}}\tau  + \frac{{\pi \Gamma {{\left( {  {\raise0.5ex\hbox{$\scriptstyle 7$}
\kern-0.1em/\kern-0.15em
\lower0.25ex\hbox{$\scriptstyle 8$}}} \right)}^2}}}{{\Gamma {{\left( {{\raise0.5ex\hbox{$\scriptstyle 3$}
\kern-0.1em/\kern-0.15em
\lower0.25ex\hbox{$\scriptstyle 8$}}} \right)}^2}}},
\end{equation}
i.e. the asymptotic dependence is $ \sim {\tau ^2}$ plus lower order correction terms\footnote{The same expression given in \citet{2014MNRAS.445.3592P}, their Eq.(23) or (A6), contained a typos that is amended here. In the RHS of their Eq.(A6) the factor 2 should read 1/2. Their Fig. (A1) is nevertheless unchanged because already plotting the correct (A6) without errors. }.  As a consequence of this, also the time averaged value $  \chi \left( \tau  \right) = \frac{1}{\tau }\int_0^\tau  {\chi \left( {\tau '} \right)d\tau '}$ will grow with the same temporal proportionality:
\begin{equation}\label{Eq037}
\begin{array}{rcl}
  \chi \left( t \right) &=& \frac{1}{\tau }\int_0^\tau  {\frac{{{{\left( {\tau \Gamma \left( {{\raise0.7ex\hbox{$3$} \!\mathord{\left/
 {\vphantom {3 8}}\right.\kern-\nulldelimiterspace}
\!\lower0.7ex\hbox{$8$}}} \right) + 2\sqrt \pi  \Gamma \left( {{\raise0.7ex\hbox{$7$} \!\mathord{\left/
 {\vphantom {7 8}}\right.\kern-\nulldelimiterspace}
\!\lower0.7ex\hbox{$8$}}} \right)} \right)}^2}}}{{4\Gamma {{\left( {{\raise0.7ex\hbox{$3$} \!\mathord{\left/
 {\vphantom {3 8}}\right.\kern-\nulldelimiterspace}
\!\lower0.7ex\hbox{$8$}}} \right)}^2}}}d\tau } \\
 &=& \frac{{{\tau ^2}}}{{12}} + \frac{{\sqrt \pi  \tau \Gamma \left( {{\raise0.7ex\hbox{$7$} \!\mathord{\left/
 {\vphantom {7 8}}\right.\kern-\nulldelimiterspace}
\!\lower0.7ex\hbox{$8$}}} \right)}}{{2\Gamma \left( {{\raise0.7ex\hbox{$3$} \!\mathord{\left/
 {\vphantom {3 8}}\right.\kern-\nulldelimiterspace}
\!\lower0.7ex\hbox{$8$}}} \right)}} + \frac{{\pi \Gamma {{\left( {{\raise0.7ex\hbox{$7$} \!\mathord{\left/
 {\vphantom {7 8}}\right.\kern-\nulldelimiterspace}
\!\lower0.7ex\hbox{$8$}}} \right)}^2}}}{{\Gamma {{\left( {{\raise0.7ex\hbox{$3$} \!\mathord{\left/
 {\vphantom {3 8}}\right.\kern-\nulldelimiterspace}
\!\lower0.7ex\hbox{$8$}}} \right)}^2}}}
\end{array}
\end{equation}
This is the equation  we are going to use below.

\subsubsection{The acceleration of convective elements}\label{appendAccel}
In $S_0$ the motion of an element of mass $m_e$ is driven by the Newton Laws ${\bm{F}}_{\rm{tot}}= {\bm{F}}_g + {\bm{F}}_P = m_e \ddot{\bm{ x}}$  where ${\bm{F}}_g$  is the gravitational force and ${\bm{F}}_P$ the force due to the pressure exerted by the surrounding medium, and the total force ${\bm{F}}_T$ is acting on the barycenter.  In$ S_1$  summing up all the contributions to the pressure  on the element surface exerted by the medium from all directions (represented by the normal  ${\bm{\hat n}}$ and the solid angle $d\Omega$)    we obtain
\begin{equation}\label{Eq032}
 - \int_{}^{} {P{\bm{\hat n}}d\Omega } = {\bm{F}}_P  =  - \left( {\frac{2}{3}\pi A\rho \xi _e^3 + \frac{4}{3}\pi g\rho \xi _e^3 + 2\pi \rho v{{\dot \xi }_e}\xi _e^2} \right),
\end{equation}
The RHS of this equation contains three terms: the  buoyancy force on the  convective element $\frac{4}{3}\pi \xi _e^3\rho {\bm{g}}$,  the inertial term of the fluid  displaced by the movement of the convective cell, i.e. the reaction mass $\frac{1}{2}\frac{4}{3}\pi \xi _e^3\rho  \equiv \frac{M}{2}$, and a new extra term $ - 2\pi \xi _e^2\rho {\bm{v}}{\dot \xi _e}$ arising from the changing  size of the convective element: the larger  the convective element, the stronger  the buoyancy effect and the larger is the velocity acquired by  the convective element. These terms must be included in the Newtonian EoM that reads\footnote{In \cite{2014MNRAS.445.3592P}, the expression for the same acceleration, their Eq. (26), contained  a typing mistake amended here. }
\begin{equation}\label{EqMot}
	{A_z} =  - g\frac{{{m_e} - M}}{{{m_e} + \frac{M}{2}}} - \frac{{2\pi \rho }}{{{m_e} + \frac{M}{2}}}v{{\dot \xi }_e}\xi _e^2.
\end{equation}
The last step now is to work out the vertical component of the acceleration $A_z$ as a function of the temperature gradient $\nabla$,  $\nabla_\mu \equiv \frac{{\partial \ln T}}{{\partial \ln \mu }}$ (gradient in molecular weight),  and $\nabla_e$ (convective element). Using the complete expression for ${A_z}$ and applying a lengthy and tedious procedure that takes into account how the densities of the medium and convective element vary with the position, one arrives to the result
\begin{equation}\label{Eq043}
{A_z} \simeq g\frac{{{\nabla _e} - \nabla  + \frac{\varphi }{\delta }{\nabla _\mu }}}{{\frac{{3{h_p}}}{{2\delta \Delta z}} + \left( {{\nabla _e} + 2\nabla  - \frac{\varphi }{{2\delta }}{\nabla _\mu }} \right)}},	
\end{equation}
with $\alpha$ and $\delta$ introduced in Sec.\ref{ion_therm} and $\varphi \equiv \frac{{\partial \ln \rho }}{{\partial \ln \mu }}$. Particularly interesting is the case of a homogeneous medium in which $\nabla_\mu$=0, in which
\begin{equation}\label{Eq044}
		A_z^\infty \simeq g\frac{{{\nabla _e} - \nabla }}{{\frac{{3{h_p}}}{{2\delta \Delta z}} + \left( {{\nabla _e} + 2\nabla } \right)}}.
	\end{equation}
If we reduced equation to the leading order in $\frac{{{h_p}}}{{\Delta r}} \to \infty $, in a chemically homogeneous convective layer we recover the well know result:
\begin{equation}\label{EqWellKnown}
		A_z \simeq  - g\frac{2}{3}\frac{\delta }{{{h_p}}}\left( {{\nabla _e} - \nabla } \right)\Delta r
\end{equation}
as asymptotic approximation of order $O\left( {\frac{A}{g}} \right)$.
We note that using Eqn.(\ref{EqWellKnown}) we can  integrate the EoM of the convective element in ${S_0}$: ${A_z} = \ddot z =  - g\frac{2}{3}\frac{\delta }{{{h_p}}}\left( {{\nabla _e} - \nabla } \right)z$. Hence, it is easy to verify that a double integration would lead $z = \frac{1}{2}{A_0}{e^{ - Xt}}\left( {{e^{2Xt}} + 1} \right)$ with ${X^2} \equiv g\frac{2}{3}\frac{\delta }{{{h_p}}}\left( {\nabla  - {\nabla _e}} \right)$ so that the velocity of the convective element will be given by $v = \dot z = X{v_0}{e^{Xt}} - \frac{1}{2}X{z_0}{e^{ - Xt}}\left( {{e^{2Xt}} + 1} \right) \simeq {X^2}t{z_0} + O{\left( t \right)^2}$. From this relation we also get ${v^2} = {A_z^2}{t^2}$. But to the leading term $\frac{{{\xi _e}}}{{{\xi _{e0}}}} \propto \frac{1}{4}\frac{{{t^2}}}{{t_0^2}}$ (see also Sec.\ref{Secref2.3}) so that ${t^2} = A_z\frac{{t_0^2}}{{{\xi _{e0}}}}$  and now remembering that ${t^2} =  - \frac{4}{A_z}{\xi _e}$,  we obtain again the proportionality ${v^2} \propto A{\xi _e}$ already presented in Sec.\ref{VElpot}.

It is then immediately evident how this expression implies the Schwarzschild criterion for convective instability (${\nabla _e} - \nabla  < 0$) as the denominator of Eq.(\ref{Eq044})  is always positive by definition. This is a very important result  because it allows us to recover the Schwarzschild and/or Ledoux criteria for instability: even with the new criterion,  the convective zones occur exactly in the same regions predicted by the Schwarzschild  criterion. For more details on this issue see  \citet[][]{2014MNRAS.445.3592P}.

\subsubsection{The final set of equation for the SFC theory}
The final step to undertake is to set up the equations  for  the convective flux, the typical dimension of the convective elements, their velocity etc. The only minor point to comment briefly concerns the possible inclusion of the conductive flux. Since from a formal point of view the conductive flux is customarily expressed in the same way as the radiative one provided the pure radiative opacity is suitably replaced by $\frac{1}{\kappa} = \frac{1}{\kappa_{{\rm{rad}}}} +\frac{1}{\kappa_{\rm{cnd}}}$ with obvious meaning of the symbols \citep[see for instance][or any other textbook]{1968pss..book.....C,1994sse..book.....K}, in the equation below we have indicated the portion of total flux carried by radiation plus conduction  with the notation $\varphi_{\rm{rad/cond}}$ and suitably redefined the  opacity $\kappa$. If conduction is not important all this reduces to the standard radiative flux. The definition of the convective flux is the standard one.

The system of equations derived by \citet[][]{2014MNRAS.445.3592P} that must  be solved to determine the convective/radiative-conductive transfer of energy in the atmosphere is:
\begin{equation}\label{MySystemApp}
\left\{ \begin{array}{rcl}
{\varphi _{{\rm{rad/cnd}}}} &=& \frac{{4ac}}{3}\frac{{{T^4}}}{{\kappa {h_p}\rho }}\nabla \\
{\varphi _{{\rm{rad/cnd}}}} + {\varphi _{{\rm{cnv}}}} &=& \frac{{4ac}}{3}\frac{{{T^4}}}{{\kappa {h_p}\rho }}{\nabla _{{\rm{rad}}}}\\
\frac{{{{  v}^2}}}{{{{  \xi }_e}}} &=& \frac{{\nabla  - {\nabla _e} - \frac{\varphi }{\delta }{\nabla _\mu }}}{{\frac{{3{h_p}}}{{2\delta {{{  v}}{t_0}\tau }}} + \left( {{\nabla _e} + 2\nabla  - \frac{\varphi }{{2\delta }}{\nabla _\mu }} \right)}}g\\
{\varphi _{{\rm{cnv}}}} &=& \frac{1}{2}\rho {c_p}T\left( {\nabla  - {\nabla _e}} \right)\frac{{{{  v}^2}{t_0}\tau }}{{{h_p}}}\\
\frac{{{\nabla _e} - {\nabla _{{\rm{ad}}}}}}{{\nabla  - {\nabla _e}}} &=& \frac{{4ac{T^3}}}{{\kappa {\rho ^2}{c_p}}}\frac{{{t_0}\tau }}{{  \xi _e^2}}\\
{{  \xi }_e} &=& {\left( {\frac{{{t_0}}}{2}} \right)^2}\frac{{\nabla  - {\nabla _e} - \frac{\varphi }{\delta }{\nabla _\mu }}}{{\frac{{3{h_p}}}{{2\delta {{{  v}}{t_0}\tau }}} + \left( {{\nabla _e} + 2\nabla  - \frac{\varphi }{{2\delta }}{\nabla _\mu }} \right)}}g  \chi \left( \tau  \right).
\end{array} \right.
\end{equation}
At each layer, this system is defined once the quantities $\left\{ {T,\kappa ,\rho ,{\nabla _{{\text{rad}}}},{\nabla _{{\text{ad}}}},\nabla _\mu,g,{c_p}} \right\}$  (considered as  averages over an infinitesimal region $dr$ \textit{and} time $dt$) are given as   input. This means that the time-scale over which these quantities vary is supposed to be much longer than the time over which the time integration of system is performed.

\end{appendix}
\label{lastpage}

\end{document}